\documentclass[journal]{IEEEtran}
\ifCLASSINFOpdf
\else
\fi

\usepackage{times,amsmath,epsfig}
\usepackage{fancyhdr}
\usepackage{amsmath}
\usepackage{amsfonts}
\usepackage{amssymb}
\usepackage[latin1]{inputenc}
\usepackage{array}
\usepackage{graphicx}
\usepackage{url}
\usepackage{bm}
\usepackage{xcolor}
\usepackage{amssymb}

\usepackage{multicol}
\usepackage{multirow}
\usepackage{amsmath,latexsym,amsbsy,amssymb,fancyhdr}
\usepackage{array}
\usepackage{enumerate}
\usepackage{amsthm}
\usepackage{booktabs} % Top and bottom rules for tables
\usepackage{multirow}
\usepackage{tabularx}
\usepackage{epsfig} %For pictures: screened artwork should be set up with an 85 or 100 line screen

\usepackage{psfrag}
\usepackage{graphicx, caption, subcaption}
\usepackage{epstopdf}
\usepackage[latin1]{inputenc}
\usepackage[colorlinks=true]{hyperref}
\hypersetup{urlcolor=blue, citecolor=blue, linkcolor=blue}

\usepackage{color}
\usepackage{cite}
\usepackage{commath}
\usepackage{mathtools}
\usepackage{systeme}

\usepackage{tabularx}
\newcolumntype{M}[1]{>{\centering\arraybackslash}m{#1}}

\usepackage{xparse}
\NewDocumentCommand{\INTERVALINNARDS}{ m m }{
	#1 {,} #2
}
\NewDocumentCommand{\interval}{ s m >{\SplitArgument{1}{,}}m m o }{
	\IfBooleanTF{#1}{
		\left#2 \INTERVALINNARDS #3 \right#4
	}{
	\IfValueTF{#5}{
		#5{#2} \INTERVALINNARDS #3 #5{#4}
	}{
	#2 \INTERVALINNARDS #3 #4
}
}
}
\usepackage{tikz, tkz-tab}

\usepackage{caption}

\newcommand{\at}[2][]{#1|_{#2}}

\newcommand{\sign}{\operatorname{sign}}

\newcommand{\R}{{\mathbb R}}

\usepackage{balance}

\begin{document}

%
% paper title
% Titles are generally capitalized except for words such as a, an, and, as,
% at, but, by, for, in, nor, of, on, or, the, to and up, which are usually
% not capitalized unless they are the first or last word of the title.
% Linebreaks \\ can be used within to get better formatting as desired.
% Do not put math or special symbols in the title.

%\title{Fundamental Dynamics and Limitations on Transmitted Power of a Two-coil Wireless Power Transfer System}
\title{Two-coil Wireless Power Transfer System Configured in Series--Series Topology: Fundamental Dynamics and Limitations on Transmitted Power}

%
% author names and IEEE memberships
% note positions of commas and nonbreaking spaces ( ~ ) LaTeX will not break
% a structure at a ~ so this keeps an author's name from being broken across
% two lines.

%\author{Zhengyu~Peng,~\IEEEmembership{Student~Member,~IEEE,}
%        Michael~Shell,~\IEEEmembership{Member,~IEEE,}
%        John~Doe,~\IEEEmembership{Fellow,~OSA,}
%        and~Jane~Doe,~\IEEEmembership{Life~Fellow,~IEEE}% <-this % stops a space
%}% <-this % stops a space

%\author{Binh~Duc~Truong,
%	Thuy~Thi-Thien~Le,
%	and~Berardi~Sensale-Rodriguez,~\IEEEmembership{Senior~Member,~IEEE}% <-this % stops a space
%}

\author{Binh~Duc~Truong,
	Thuy~Thi-Thien~Le,
	and~Berardi~Sensale-Rodriguez
}

% The paper headers

%\markboth{IEEE Transactions on Circuits and Systems I: Regular Papers}
%{B. Truong \MakeLowercase{\textit{et al.}}: Fundamental dynamics and limitations on the transmitted power of a two-coil WPTS}

%\markboth{Fundamental dynamics - Frequency splitting suppression - Limitations on power}
%{B. Truong \MakeLowercase{\textit{et al.}}: Fundamental dynamics and limitations on the transmitted power of a two-coil WPTS}

%\twocolumn[\begin{@twocolumnfalse}
  
% make the title area
\maketitle

% As a general rule, do not put math, special symbols or citations
% in the abstract or keywords.

\begin{abstract}
	
The dynamics and performance of a two-coil resonant coupled wireless power transfer system (WPTS) are investigated. At high coupling coefficient between the two resonators, the frequency splitting phenomenon occurs, in which the power transferred to the load attains its maximum at two frequencies that are away from the resonance frequency. However, this behavior is not a universal property; there exist certain regions of resonator intrinsic parameters at which it is not present for any coupling strength. Therefore, in order to suppress such a phenomenon, there is no need to constrain the coupling of the transmitter and receiver below a certain level as widely reported in the literature. For low-power wearable and implantable applications or wireless sensor nodes, optimizing the power delivered to the load is essential. Based on a general maximum power transfer theorem, we derive a rigorous asymptotic upper bound for the power that can be delivered to an arbitrary load from a generic source. Our results also quantitatively reveal the essential role of the unloaded quality factors of the two resonators, the coupling between them, and their direct impacts on the actual output power. Furthermore, we discuss that in contrast to the often employed operating power gain, the transducer power gain constitutes a more suitable metric to describe (and optimize) system efficiency. In general, when the source and load impedances are predetermined, maximum transferable power can be reached by using bi-conjugate impedance matching circuits. Once the transferred power reaches its physical bound, the transducer power gain and the operating power gain collapse to a unique global optimal solution. For the completeness of the analysis, a distinctive dynamics of the perfect coupled system, where the coupling coefficient equals unity, is discussed. In this circumstance, it is shown that the output power tends to saturate at a frequency-independent constant value when the driving frequency is relatively large (approaches infinity).

%The dynamics and performance of a two-coil resonant coupled wireless power transfer system are investigated. At high coupling, the frequency-splitting phenomenon occurs, in which the power transferred to the load attains its maximum at two frequencies away from the resonance frequency. However, this behavior is not a universal property; there exist certain regions of resonator intrinsic parameters in which it is not present for any coupling strength. Therefore, in order to suppress such a phenomenon, there is no need to constrain the coupling of the transmitter and receiver below a certain level as widely reported in the literature. For low-power applications, optimizing the received power is essential. We derive a rigorous asymptotic upper bound for the power that can be delivered to an arbitrary load from a generic source. Our results quantitatively reveal the direct impacts of the unloaded $Q$--factors of the two resonators and the coupling between them on the actual output power. We discuss that, in contrast to the often employed operating power gain, the transducer power gain constitutes a more suitable metric to optimize system efficiency. Once the transferred power reaches its physical bound, the two gains collapse to a unique global optimal solution.

\end{abstract}

% Note that keywords are not normally used for peerreview papers.
\begin{IEEEkeywords}
Fundamental dynamics, Frequency-splitting phenomenon, Wireless Power Transfer, Limitation on Transmitter Power, Series-Series Configuration.
\end{IEEEkeywords}

%\end{@twocolumnfalse}]

% Put footnotes here
{
  \renewcommand{\thefootnote}{}%
  \footnotetext[1]{B. Truong, T. Le and B. Sensale-Rodriguez are with the Department of Mechanical Engineering, Department of Mathematics and Department of Electrical and Computer Engineering, respectively, University of Utah, Salt Lake City, UT, USA. E-mail: Binh.D.Truong@utah.edu
%  	Email: Binh.D.Truong@utah.edu and berardi.sensale@utah.edu
}
%  \footnotetext[2]{}
}
 
% For peer review papers, you can put extra information on the cover
% page as needed:
% \ifCLASSOPTIONpeerreview
% \begin{center} \bfseries EDICS Category: 3-BBND \end{center}
% \fi
%
% For peerreview papers, this IEEEtran command inserts a page break and
% creates the second title. It will be ignored for other modes. 
\IEEEpeerreviewmaketitle

\section{Introduction}

\IEEEPARstart{W}{ireless} power transfer is an effective means for charging consumer electronic devices such as wireless sensor nodes in machinery and on human body \cite{Agarwal2017, Kim2017}. The feasibility of a wireless power transfer system (WPTS) utilizing magnetically coupled resonators was demonstrated both theoretically and experimentally \cite{Sample2011,Kurs2007}. The wireless power transfer mechanism in this configuration has been extensively investigated in the literature based on either coupled-mode theory \cite{Kurs2007} or equivalent circuit theory \cite{Kiani2012b}. The mechanism behind the transmission process can be also thoroughly explained in the electromagnetic domain \cite{Guo2017}. These studies so far reveal a frequency splitting phenomenon occurring when the resonators are over-coupled, in which the power generated at the load reaches its peaks not at the original resonance frequency but at two adjacent frequencies. Although it has been widely studied over the last decade, exact analytical solutions describing the position of these maxima and minimum as well as fundamental conditions defining the presence of this phenomenon are not available in the literature and therefore are of great interest.

Frequency splitting phenomena require the practical WPTS to either track the optimal frequency \cite{Assawaworrarit2017, Sample2011, Heebl2014} or vary the distance between the transmitter and receiver \cite{Zhang2014} in order to maintain a sufficient power delivered to a load. Several authors have attempted to eliminate the influences of the frequency splitting behavior by using tunable impedance matching networks \cite{Heebl2014, Luo2014} or through improving designs of the resonance coils \cite{Ettorre2012, Lee2013, Lyu2015, Zhang2017}. All of these analyses were specific to those particular systems. A generalized solution for frequency splitting suppression, which is independent of system structure, has not been fully developed yet.

A well-known technique to analyze the frequency splitting characteristics is to consider the input impedance, $Z_\mathrm{in}$. The splitting frequencies are assumed to coincide with the resonance frequencies that are determined by setting the imaginary part of $Z_\mathrm{in}$ equal to zero, $\Im\{Z_\mathrm{in}\} = 0$. We note that the resonance frequencies are also referred to as zero-phase angle or bifurcation frequencies. Following that, the frequency splitting appears if the coupling coefficient $k$ of the two resonators goes beyond a critical value $k_\mathrm{c}$ characterized by the condition such that there exist three resonance frequencies. Therefore, in order to avoid the frequency splitting, $k$ should be selected to be less than $k_\mathrm{c}$, which in practice can be done by restricting the minimum distance between the transmitter and receiver. This method has been extensively used even in the most recent works \cite{Abdelatty2019, Aditya2019}. However, we believe that the exact solutions of the splitting frequencies must be obtained by solving the equation $\dif P_\mathrm{L}/\dif \omega = 0$ instead of $\Im\{Z_\mathrm{in}\} = 0$, where $P_\mathrm{L}$ is the power transferred to the load and $\omega$ is the driving angular frequency. This different point of view could shed new light on the frequency splitting problem.

Most of the investigations on WPTSs are focused on the operating power gain (OPG), which is defined by the ratio between the power transferred to the load and the power input to the network \cite{Aditya2019, Aldhaher2018}. The OPG is referred to as the link efficiency $\eta$ in the literature. The input power is dependent on the input impedance of the system and is a function of the resonator characteristics, the load resistance, the coupling strength and the driving frequency. It changes with variation of any of these parameters. However, optimizing OPG does not always ensure maximum output power delivered to the load \cite{TruongWPTS2018} and thus may not be the key factor for many low-power applications, such as biomedical sensing systems. The studies in \cite{Aditya2019, Aldhaher2018} do not take into account the role of the power available from the source, $P_\mathrm{avs}$, which can be considered as the strict upper-bound on the output power of a WPTS. For a realistic device, it is essential to determine the maximum transferable power of an arbitrarily given WPTS subject to mild restrictions on the physical parameters. Different from other published works, we treat the actual transferred power $P_\mathrm{L}$ as the central objective of the investigation rather than the transfer efficiency $\eta$. From this distinguished perspective, we expect to provide an additional understanding of the performance of the WPTS.

The frequency splitting issue was partially discussed in \cite{TruongTIE2020}, in which the problem was formulated for a system with matched resonance frequencies. Several possible scenarios were addressed, in which two dimensionless parameters -- the loaded $Q$--factors of the two resonators -- were chosen as the objectives of the study. However, analytical solutions to those equations for a particular dimensional parameter such as capacitance have not been found yet. Moreover, the relationship between the resonance and splitting frequencies under different conditions has not been fully exploited and is worthwhile to clarify. It is essential to emphasize that the frequency splitting phenomenon can occur even when the resonance frequencies of the transmitter and receiver are not identical. Therefore, a more thorough analysis and a method to solve the general case with arbitrary parameters of the two resonators are of great interest to further explore.

In this work, we theoretically investigate and tackle all the questions raised above with two major issues, (i) frequency splitting and (ii) an upper bound on power transferred. We focus on (but not entirely limited to) a two-coil WPTS configured in series, which is perhaps the most widely used in practice. In particular, we provide a universal comprehensive picture of the fundamental dynamics of the considered system. We propose an efficient approach to prevent the frequency splitting property regardless of how the two coils are designed and how strong the coupling between them is. A simple implementation is to adjust the nominal resonance frequency by changing the added capacitance $C_{1/2}$; an explicit solution of $C_{1/2}$ is given. We also analytically determine the asymptotic rigorous upper bound on the optimum achievable output power, showing a general framework for designing and optimizing the system performance. Several case studies are experimentally validated to support and demonstrate our theoretical hypothesis. Note that, although the main target of the paper is low-power applications, all presented arguments hold true independently of any particular devices.

The article is organized as follows. First of all, in Section \ref{Model}, we introduce an equivalent circuit model of a two-coil WPTS and establish the analytical solution of the power transferred to the load. Section \ref{FreqSplit} is devoted to investigating the frequency splitting phenomenon and the performance of the output power with respect to the drive frequency. Then, Section \ref{FreqsplElim} focuses on examining the general conditions under which the frequency splitting behavior is suppressed for any coupling strength.
In Section \ref{ResFreqsVsExtrFreqs}, a counterexample is provided to prove that there are scenarios where the delivered power has a unique maximum while exhibiting three distinguished resonance frequencies. In addition, we also show that the frequency splitting phenomenon can occur even when there only exists a single zero-phase angle frequency. Therefore, studying the imaginary part of the input impedance is not sufficient to describe the physical insights of the frequency splitting characteristics. These findings return the problem to its true nature and essence. The principle of the power optimization problem is presented in Section \ref{PowerOpt}. A unique asymptotic property of a perfect coupled system is discussed in Section \ref{UnityCoupling}. In the last section, the experimental validations of some essential findings are carried out.

\section{Equivalent circuit model} \label{Model}

%\begin{figure}[!t] %[!thb]
%	\centering
%	\includegraphics[width=0.425\textwidth]{WPTtech.pdf}
%	\caption{\small Inductive-based wireless power transfer concept.}
%	\label{Fig:concept}
%\end{figure}
%Figure \ref{Fig:concept} depicts the key concept of 
In an inductive-based WPTS, a transmitter $L_1$ driven by an electric power source generates a time-varying electromagnetic field traveling through space, or some other medium, across a receiver $L_2$ where the electromagnetic energy is extracted and then supplied to an electrical load. The magnetic induction between the transmitter and receiver is modeled by a mutual inductance $M$, which depends on the two coil geometries and their separation distance. A widely dimensionless figure of merit is the coupling coefficient defined by $k = M/\sqrt{L_1 L_2}$, which represents the fraction of magnetic flux density generated by $L_1$ that passes through $L_2$ when $L_2$ is in open circuit condition \cite{Pierce1907}. Resonant inductive coupling (or strongly coupled magnetic resonance) WPTSs are one of the most commonly used structures in which each transmitting/receiving coil is connected with a tunable capacitor in order to form a resonator. 

\begin{figure}[!t]%[!thb]
	\centering
	\includegraphics[width=0.385\textwidth]{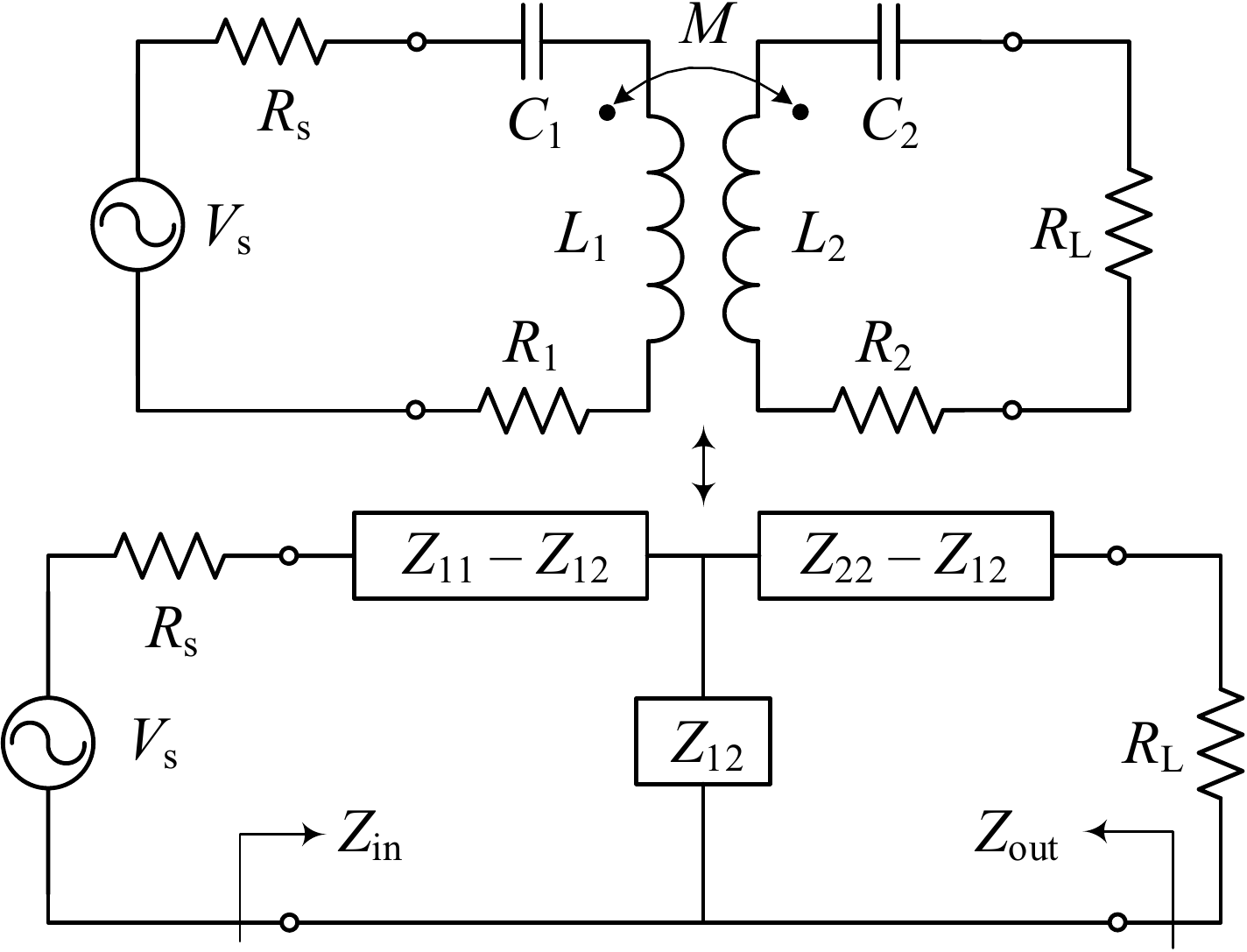}
	\caption{\small Equivalent circuit model of two-coil WPTS configured in series and its representation in terms of Z-parameters.}
	\label{Fig:2coilSys}
\end{figure}
Figure \ref{Fig:2coilSys} shows an equivalent circuit model of a two-coil series-series compensated resonant WPTS, where $(L_1R_1C_1)$ and $(L_2R_2C_2)$ are the inductance, parasitic resistance and added capacitance of the transmitter and receiver coils, respectively. The source is represented by a series circuit with a source voltage $V_\mathrm{s}$ and a resistor $R_\mathrm{s}$; $R_\mathrm{L}$ is the load resistance. The parasitic capacitances of the two coils are considered to be small and therefore neglected. This model was verified and used in the literature (see \cite{Sample2011, Kiani2012b} for example), which is reliable for further analysis.
The T-network equivalent-circuit impedances of the inductive link are
%\begin{align}
%\small
%Z_{11} &= R_1 + j \Big(\omega L_1 - \frac{1}{\omega C_1}\Big), \label{Z11} \\
%Z_{22} &= R_2 + j \Big(\omega L_2 - \frac{1}{\omega C_2}\Big), \\
%Z_{12} &= Z_{21} = j\omega M \label{Z12}
%\end{align}
\begin{align}
\small
Z_{11} &= R_1 + j \big(\omega L_1 - 1/(\omega C_1)\big), \label{Z11} \\
Z_{22} &= R_2 + j \big(\omega L_2 - 1/(\omega C_2)\big), \\
Z_{12} &= Z_{21} = j\omega M \label{Z12}
\end{align}
where $\omega$ is the angular driving frequency \cite{TruongWPTS2018}.
Therefore,
%\begin{align}
%\small
%Z_\mathrm{in} &= R_1 + j \Big(\omega L_1 - \frac{1}{\omega C_1} \Big) + \frac{(\omega M)^2}{R_2 + \displaystyle j \Big(\omega L_2 - \frac{1}{\omega C_2} \Big) + R_\mathrm{L}}, \label{Z_in} \\
%Z_\mathrm{out} &= R_2 + j \Big(\omega L_2 - \frac{1}{\omega C_2} \Big) + \frac{(\omega M)^2}{R_1 + \displaystyle j \Big(\omega L_1 - \frac{1}{\omega C_1} \Big) + R_\mathrm{s}}. \label{Z_out}
%\end{align}
\begin{align}
\small
\begin{split}
Z_\mathrm{in} &= R_1 + j \big(\omega L_1 - 1/(\omega C_1) \big) \\
& + (\omega M)^2 \Big{/} \big[ R_2 + \displaystyle j \big(\omega L_2 - 1/(\omega C_2) \big) + R_\mathrm{L} \big], \label{Z_in}
\end{split}
\\
\begin{split}
Z_\mathrm{out} &= R_2 + j \big(\omega L_2 - 1/(\omega C_2) \big) \\
& + (\omega M)^2 \Big{/} \big[ R_1 + \displaystyle j \big(\omega L_1 - 1/(\omega C_1) \big) + R_\mathrm{s} \big]. \label{Z_out}
\end{split}
\end{align}
The general explicit form of the output power is derived as follows \cite{TruongWPTS2018}
%\begin{align}
%\small
%\begin{split}
%P_\mathrm{L} &= \frac{1}{2} \abs{V_\mathrm{s}}^2 \frac{\abs{Z_{21}}^2 R_\mathrm{L}}{\abs{(Z_{11} + Z_\mathrm{s}) (Z_\mathrm{out} + Z_\mathrm{L})}^2} \\
%&= \frac{1}{2} \frac{\abs{V_\mathrm{s}}^2 (\omega M)^2 R_\mathrm{L} \kappa }{\displaystyle \Big[ \kappa (R_1 + R_\mathrm{s}) (\omega M)^2 + R_\mathrm{L} + R_2 \Big]^2 + \Big[ \omega L_2 - \frac{1}{\omega C_2} - (\omega M)^2 \Big(\omega L_1 - \frac{1}{\omega C_1} \Big) \kappa \Big]^2} \label{PL2coilsGen} %\\
%\end{split}
%\end{align}
\begin{align}
\small
\begin{split}
& P_\mathrm{L} = \frac{1}{2} \abs{V_\mathrm{s}}^2 \frac{\abs{Z_{21}}^2 R_\mathrm{L}}{\abs{(Z_{11} + Z_\mathrm{s}) (Z_\mathrm{out} + Z_\mathrm{L})}^2} \\
&= \frac{1}{2} \abs{V_\mathrm{s}}^2 (\omega M)^2 R_\mathrm{L} \kappa \Big{/} \! \left\{ \displaystyle \big[ \kappa (R_1  + R_\mathrm{s}) (\omega M)^2 + R_\mathrm{L} + R_2 \big]^2 \right. \\
& \left. + \big[ \omega L_2 - 1/(\omega C_2) - (\omega M)^2 \big(\omega L_1 - 1/(\omega C_1) \big) \kappa \big]^2 \right\}
\label{PL2coilsGen}
\end{split} \\
%\end{align}
%where
%\begin{align}
%\small
&\text{where } \kappa = \big[ \big(\omega L_1 - 1/(\omega C_1) \big)^2 + (R_1 + R_\mathrm{s})^2 \big]^{-1}. \label{kappa}
\end{align}
Formula \eqref{PL2coilsGen} is the central objective throughout the paper.

In this work, we choose to utilize two-port theory to investigate the performance of the two-coil WPTS since it is more familiar to electrical and electronics engineers than coupled mode theory, and therefore may be a better choice to bridge the gap between physics and engineering. Another advantage of the former approach is its ability to express the coupled system in explicit algebraic forms in the frequency domain instead of describing the physical coupling by differential equations. The equivalence of coupled mode and circuit theory was analyzed in, e.g., \cite{Kino1962, Kiani2012b, Ricketts2013}.

\section{The frequency splitting phenomenon and fundamental dynamics of the series-configured Wireless Power Transfer System} \label{FreqSplit}

In this section, we examine the changes on the transferred power \eqref{PL2coilsGen} with respect to the driving angular frequency and reveal fundamental dynamics of the two-coil WPTS in series-series configuration. We consider a general case for an arbitrary set of parameters, regardless of matched or unmatched resonance conditions.

The stationary point(s) are determined by equation $\dif P_\mathrm{L}/\dif\omega = 0$, which can be written as
\begin{align} 
\small
f(\Omega) & = \alpha \Omega^4 + \beta \Omega^2 + \gamma \Omega + \lambda = 0. \label{fOmegaEq}
\end{align}
The coefficients $\alpha, \, \beta, \, \gamma$ and $\lambda$ can be normalized and expressed as functions of architecture-independent parameters, such as the electrical time constants $(\tau_\mathrm{s}, \, \tau_\mathrm{L})$ and the angular resonance frequencies $(\omega_1, \, \omega_2)$ of the transmitter and receiver coils, as follows
\begin{align*}
\small
\alpha &= -\frac{(k^2 - 1)^2}{(\omega_1 \omega_2)^4}, \\
\begin{split}
\beta &= \frac{1}{(\omega_1 \omega_2)^4} \bigg[ \frac{1}{(\tau_\mathrm{s} \tau_\mathrm{L})^2} + 2(1-k^2) (\omega_1 \omega_2)^2 \\
& + (\omega_1^2 + \omega_2^2)^2 - 2 \Big(\frac{\omega_2}{\tau_\mathrm{s}} \Big)^2 - 2 \Big(\frac{\omega_1}{\tau_\mathrm{L}} \Big)^2 \bigg], 
\end{split} \\
%\begin{split}
%\gamma &= \frac{2}{(\omega_1 \omega_2)^4} \bigg[ \Big(\frac{\omega_1^2}{\tau_\mathrm{L}} \Big)^2 + \Big(\frac{\omega_2^2}{\tau_\mathrm{s}} \Big)^2 \\
%& - 2 (\omega_1 \omega_2)^2 (\omega_1^2 + \omega_2^2) \bigg], 
%\end{split} \\
\gamma &= \frac{2}{(\omega_1 \omega_2)^4} \bigg[ \Big(\frac{\omega_1^2}{\tau_\mathrm{L}} \Big)^2 \!+\! \Big(\frac{\omega_2^2}{\tau_\mathrm{s}} \Big)^2 
\!-\! 2 (\omega_1 \omega_2)^2 (\omega_1^2 \!+\! \omega_2^2) \bigg], \\
\lambda &= 3, \\
\tau_\mathrm{s} &= L_1/(R_1 + R_\mathrm{s}), \,
\tau_\mathrm{L} = L_2/(R_2 + R_\mathrm{L}), \\
\omega_1 &= 1/\sqrt{L_1 C_1}, \,
\omega_2 = 1/\sqrt{L_2 C_2}.
\end{align*}
The function $f(\Omega)$ is a real coefficient polynomial of degree four without third degree term, in which all physical parameters such as resistance, inductance and capacitance are positive. Without loss of generality, we assume that $k^2 \neq 1$ (otherwise, the quartic polynomial $f(\Omega)$ reduces to a quadratic function), therefore $\alpha < 0$. 

By introducing intermediate parameters $p = \beta/\alpha, \,\, q = \gamma/\alpha$ and $r = \lambda/\alpha$, we can write \eqref{fOmegaEq} in the standard form
\begin{align} 
\small
f_\mathrm{s}(\Omega) = \Omega^4 + p \Omega^2 + q \Omega + r = 0. \label{FOmegaStand}
\end{align}
$f_\mathrm{s}(\Omega)$ collapses to the form shown in \cite{TruongTIE2020} when the two resonance frequencies are identical, $\omega_1 = \omega_2$. The cubic resolvent of the quartic equation \eqref{FOmegaStand} is defined as \cite{Zwillinger2011, Janson2010}
\begin{align} 
\label{Resolvent}
\small
t^3 - p t^2 - 4 r t + (4 p r - q^2) = 0.
\end{align}
As proven in \cite{Janson2010}, a third degree polynomial with real coefficients always has at least one real root. Let $u$ be a real root of \eqref{Resolvent} (a complete set of solutions can be found in \cite{Cardano2007} and was presented in Appendix A of our previous paper \cite{TruongTIE2020}), then the four roots of the original quartic \eqref{fOmegaEq} are given by \cite{Zwillinger2011, Janson2010}
\begin{align} 
\small
\Omega_1 &= (G + H)/2, \label{GenSol1} \\
\Omega_2 &= (G - H)/2, \label{GenSol2} \\
\Omega_3 &= (-G + I)/2, \label{GenSol3} \\
\Omega_4 &= (-G - I)/2 \label{GenSol4}
\end{align}
\begin{align*}
\text{where }
\small
G &= \sqrt{u - p}, \\
H &= \begin{cases}
\displaystyle \sqrt{-G^2 - 2 \big(p + q/G \big) }, & G \neq 0 \\
\displaystyle \sqrt{- 2p + 2\sqrt{u^2 - 4 r} }, & G = 0
\end{cases} , \\
I &= \begin{cases}
\displaystyle \sqrt{-G^2 - 2 \big(p - q/G \big) }, & G \neq 0 \\
\displaystyle \sqrt{- 2p - 2\sqrt{u^2 - 4 r} }, & G = 0 
%\label{InterParam_I}
\end{cases}.
\end{align*}

It should be noted that the notation in [\eqref{GenSol1}-\eqref{GenSol4}] is the same as in [(15--(18)] in \cite{TruongTIE2020}. However, [\eqref{GenSol1}-\eqref{GenSol4}] are \textit{the exact solutions for the general case where the transmitter and receiver parameters are arbitrary}. On the contrary, the corresponding formulas derived in \cite{TruongTIE2020} are only applicable when the resonance frequencies are matched.
In \cite{TruongTIE2020}, we showed that equation $\dif P_\mathrm{L}/\dif\omega = 0$ (or equivalently, equations \eqref{fOmegaEq} and \eqref{FOmegaStand}) must have at least one positive solution.
We now note the facts that, (i) there must exist an interval $(0, \, \epsilon_1)$ for some $\epsilon_1 > 0$ on which $P_\mathrm{L}$ increases and an interval $(\epsilon_2, \, +\infty)$ for some $\epsilon_2 > 0$ on which $P_\mathrm{L}$ decreases, and (ii) \eqref{FOmegaStand} has at most four solutions [\eqref{GenSol1}-\eqref{GenSol4}]. Thus, equation $\dif P_\mathrm{L}/\dif\omega = 0$ can only have either one or three positive solution(s); if it has two or four positive solutions, the conditions in (i) are violated.

\textit{As a summary, there are only two fundamental behaviors of $P_\mathrm{L}$ with respect to $\omega$: the output power has either a unique maximum or two maxima and one minimum}. No other possibilities exist. This is a unique property of the two-coil configuration in comparison with three- or four-coil systems, in which more than two power peaks can be observed under the right circumstances.
The latter results in the well-known frequency splitting phenomenon that has been experimentally reported in several physical systems. 
A visual summary of these two characteristics are shown in Appendix \ref{SignVariation}, where signs of $\mathrm{d}P_\mathrm{L}/\mathrm{d} \omega$ and variations of $P_\mathrm{L}$ are provided in detail. 
All these findings complete the analysis of the dynamic performance of the series-configured WPTS that we partially discussed in \cite{TruongWPTS2018}.

Let us consider the necessary condition(s) such that \eqref{FOmegaStand} has three distinguishable positive solutions and where therefore the frequency splitting takes place. Firstly, $G, \,\, H$ and $I$ must be real and non-negative. If either of them is complex and not real, \eqref{FOmegaStand} only has at most two real solutions. In the case $\{G, \, H, \, I\} \in \R_{\geq 0}$ and \eqref{FOmegaStand} has four real solutions, since  $\Omega_4 \leq 0$, the three possible non-negative solutions are $\{\Omega_1,\, \Omega_2,\, \Omega_3\}$. Then accordingly, the extreme frequencies at which $P_\mathrm{L}$ obtains its maximum and minimum values are $^\mathrm{i}\omega_\mathrm{e} = \sqrt{\Omega_\mathrm{i}}$ where $\mathrm{i}\in \{1, \,2, \,3 \}$. However, $\Omega = 0$ is not a solution of equation \eqref{FOmegaStand}, hence $G, \, H$ and $I$ are not simultaneously 0 and $\{I \neq G, \, G \neq H \}$. In addition, $\{\Omega_1,\, \Omega_2,\, \Omega_3\}$ are different, therefore the complete set of necessary conditions is \{$I > G > H > 0$ and $(I \pm H) \neq 2 G$\}. The frequency splitting is not present if any of these conditions is not fulfilled.

%\small
%\begin{table}[!t]
%	\centering
%	\caption{\small An exemplary set of system parameters adapted from \cite{Assawaworrarit2017}.}%
%	%	\medskip
%	%	\resizebox{0.6\columnwidth}{!}{
%	\begin{tabular}{l l l l} \toprule[1.0pt]
%		%		\hline
%		\multicolumn{2}{c} {\textbf{Transmitter}} & \multicolumn{2}{c} {\textbf{Receiver}}  \\
%		%heading
%		%		\hline 
%		\midrule[0.5pt]
%		$C_1$, pF & 444 & $C_2$, pF & 454 \\
%		$L_1$, $\mu$H & 9.13 & $L_2$, $\mu$H & 8.92 \\
%		$R_1$, $\Omega$ & 2.25 & $R_2$, $\Omega$ & 2.50  \\
%		\midrule[0.5pt]
%		& {\textbf{Source}}	& $R_\mathrm{s}$, $\Omega$ & 50	\\
%		\bottomrule[1.0pt]
%	\end{tabular}%
%	%	}
%	\label{Tab:SysParamters} % is used to refer this table in the text
%\end{table}
%\normalsize

\section{General conditions under which the frequency splitting phenomenon does not occur.} \label{FreqsplElim}

We have shown that when the frequency splitting behavior occurs, there are exactly two local maxima and one minimum. In general, this phenomenon is not a universal property of a WPTS. The aim of this section is to point out scenarios in which \eqref{fOmegaEq} has a unique positive root, meaning that the frequency splitting phenomenon is eliminated \textit{for any coupling strength}. One approach is to negate the necessary conditions shown in Section \ref{FreqSplit} and solve for a chosen variable. However, we realize that it leads to cumbersome relations among system parameters which are difficult to interpret further. Thus, we propose a more efficient alternative method below.

We proved that \eqref{fOmegaEq} must have at least one positive solution \cite{TruongTIE2020}. Therefore, if \eqref{fOmegaEq} has at most one positive solution in a set of system parameters, as a consequence, it is simultaneously the unique positive solution in that set. We now apply the Descartes' sign rule to determine the maximum number of positive real roots of a polynomial, denoted by $n$. Here, $n$ is the number of sign changes, as we proceed from the lowest to the highest order (ignoring orders that do not appear). Considering the polynomial $f(\Omega)$, since $\alpha < 0$ and $\lambda > 0$, $n = 1$ \textit{if either of the following conditions is satisfied}: (i) $\beta \leq 0$, (ii) $\gamma \geq 0$. The complete Descartes' sign rule table is shown in Appendix \ref{Descartes}. 
The sign of a real number does not change if we divide or multiply it by another positive number. Letting $\beta_\mathrm{s} = \beta (\omega_1 \omega_2)^2/2$ and $\gamma_\mathrm{s} = \gamma \omega_2^2 /2 $, we obtain $\sign(\beta) = \sign(\beta_\mathrm{s})$ and $\sign(\gamma) = \sign(\gamma_\mathrm{s})$. The parameters $\beta_\mathrm{s}$ and $\gamma_\mathrm{s}$ are given by dimensionless forms as follows
\begin{align*}
\small
\beta_\mathrm{s} & \!=\! 1 \! + \! \frac{1}{2} \frac{1}{(Q_\mathrm{s} Q_\mathrm{L})^2} \! + \! \frac{1}{2} \big( \frac{\omega_1}{\omega_2} \! + \! \frac{\omega_2}{\omega_1} \big)^2 \! - \! \big( \frac{1}{Q_\mathrm{s}^2} \! + \! \frac{1}{Q_\mathrm{L}^2} \big) \! - \! k^2, \\
\gamma_\mathrm{s} &= \big( \frac{\omega_2}{\omega_1} \big)^2 \frac{1}{Q_\mathrm{s}^2} + \frac{1}{Q_\mathrm{L}^2} - 2 \Big[1 + \big( \frac{\omega_2}{\omega_1} \big)^2 \Big]
\end{align*}
where the loaded quality factors of the two coils are
%\begin{align}
%\small
%Q_\mathrm{s} &= \frac{\omega_1 L_1}{R_1 + R_\mathrm{s}} = \omega_1 \tau_\mathrm{s}, \label{Qs_at_omega1} \\
%Q_\mathrm{L} &= \frac{\omega_2 L_2}{R_2 + R_\mathrm{L}} = \omega_2 \tau_\mathrm{L}. \label{QL_at_omega2}
%\end{align}
\begin{align}
\small
Q_\mathrm{s} = \frac{\omega_1 L_1}{R_1 + R_\mathrm{s}} = \omega_1 \tau_\mathrm{s}, \label{Qs_at_omega1} \,
Q_\mathrm{L} = \frac{\omega_2 L_2}{R_2 + R_\mathrm{L}} = \omega_2 \tau_\mathrm{L}.
\end{align}
The two conditions (i) and (ii) now become (I) $\beta_\mathrm{s} \leq 0$ and (II) $\gamma_\mathrm{s} \geq 0$ respectively.

\textit{This analysis reveals that, in addition to the coupling coefficient, the loaded quality factors and the resonance frequencies of the transmitter and the receiver are decisive parameters that directly affect the existence of the frequency splitting phenomenon.} 

If $(k, \, Q_\mathrm{s}, \, Q_\mathrm{L}, \, \omega_{1}, \, \omega_{2})$ satisfy (I) or (II), the frequency splitting behavior is not present. In particular, for a given coupling coefficient, lower resonance frequencies (and therefore lower coil quality factors) lead to suppression of this dynamical phenomenon. We can utilize (I) and (II) as a general framework to determine relations between the key parameters of the two resonators in order to eliminate such a phenomenon.

An example is presented below. We choose to treat the added capacitors $C_\mathrm{i}$ as dependent variables instead of the coil inductances $L_\mathrm{i}$ or parasitic resistances $R_\mathrm{i}$ ($i = \{1, \, 2\}$). This would also seem to be more common from a practical standpoint since $C_\mathrm{i}$ can be adjusted easier than other parameters. To reduce the complexity of the problem, we consider the case in which the two resonators are identical, meaning that $L_\mathrm{i} = L, \,\, C_\mathrm{i} = C, \,\, R_\mathrm{i} = R$ and $\omega_\mathrm{i} = \omega_0 = 1/\sqrt{LC}$.

The expressions of $\beta_\mathrm{s}$ and $\gamma_\mathrm{s}$ reduce to
\begin{align*}
\small
\beta_\mathrm{s} &= 3 +  \frac{1}{2} \frac{1}{(Q_\mathrm{s} Q_\mathrm{L})^2} - \Big( \frac{1}{Q_\mathrm{s}^2} + \frac{1}{Q_\mathrm{L}^2} \Big) - k^2, \\
\gamma_\mathrm{s} &= \frac{1}{Q_\mathrm{s}^2} + \frac{1}{Q_\mathrm{L}^2} - 4.
\end{align*}
The inequality equations are then solved for $C > 0$.

\textbf{Case (I):}
\begin{align*}
\small
\beta_\mathrm{s} & \leq 0 \Longleftrightarrow
\begin{cases}
\mu \geq 0 \\
\displaystyle C \in \mathbb{R}_1 = [\max\{0, \, \xi_1\}, \,\xi_2] \setminus \{0\} 
\end{cases}\\
\text{where }
\sigma_1 &= (R + R_\mathrm{s})^2 (R + R_\mathrm{L})^2 > 0, \\
\sigma_2 &= (R + R_\mathrm{s})^2 + (R + R_\mathrm{L})^2 > 0, \\
\mu &= \sigma_2 ^2 - 2 (3-k^2) \sigma_1,\\
\xi_1 &= \frac{L}{\sigma_1} (\sigma_2 - \sqrt{\mu}) \leq  \xi_2, \\
\xi_2 &= \frac{L }{\sigma_1} (\sigma_2 + \sqrt{\mu}) > 0.
\end{align*}

\textbf{Case (II):}
\begin{align*}
\small
%\begin{split}
%\gamma_\mathrm{s} \geq 0 \iff C \in \mathbb{R}_2 = \interval[{\, &C_\gamma, +\infty }) \\
%\text{where } &C_{\gamma} = \frac{4 L}{\sigma_2}.
%\end{split}
\gamma_\mathrm{s} \geq 0 \iff C \in \mathbb{R}_2 = \interval[{\, C_\gamma, +\infty }) 
\text{ where } C_{\gamma} = \frac{4 L}{\sigma_2}.
\end{align*}
\textit{Then the compact-form solution of $C$ is }
\begin{align} \label{CSols}
\small
\begin{cases} 
C \in \mathbb{R}_{\mu-} = \varnothing \cup \mathbb{R}_2 = \mathbb{R}_2, & \text{ if } \mu < 0 \\
C \in \mathbb{R}_{\mu+} = \mathbb{R}_1 \cap \mathbb{R}_2, & \text{ if } \mu \geq 0
\end{cases} .
\end{align}

%{
%\small
\begin{table*}[!t]%[!thb]
	\centering
	\caption{\small Solution of $C$ such that the frequency splitting phenomenon does not occur.}%
	\begin{tabular}{ c c c p{0.75cm} c c c } % centered columns (4 columns)
		%		\hline\hline
		\toprule %[1.0pt]
		\centering
		& \multicolumn{2}{c}{$\xi_1 \leq 0$} &  &\multicolumn{3}{c}{$\xi_1 > 0$} \\
		\midrule[0.5pt]
		\multirow{2}{*}{$\mu \geq 0$} & $C_\gamma \leq \xi_2 $ & $\xi_2 < C_\gamma $ & & $C_\gamma \leq \xi_1 \leq \xi_2$ & $\xi_1 < C_\gamma \leq \xi_2$ &  $\xi_2 < C_\gamma $ \\
		%		\midrule[0.5pt]
		%		\hline
		\cmidrule{2-7}
		& $\interval({0,+\infty})$ & $\interval({0, \,\, \xi_2}] \cup \interval[{C_\gamma, +\infty })$ &  &$\interval[{C_\gamma, +\infty })$ & $\interval[{\xi_1, +\infty})$ & $\interval[{\xi_1, \,\, \xi_2}] \cup \interval[{C_\gamma, +\infty })$ \\
		\midrule[0.5pt]
		$\mu < 0$ & \multicolumn{5}{c}{$\interval[{C_\gamma, +\infty })$} \\
		\bottomrule[1.0pt]
	\end{tabular}
	\label{Table:C_solution_Sum} % is used to refer this table in the text
\end{table*}
%}

We see that, for any values of $0 \leq k^2 < 1$, there always exists $C$ in one or more region(s) (defined by other parameters) in which $P_\mathrm{L}$ has a unique stationary point.
The solution of $C$ is summarized in Table \ref{Table:C_solution_Sum}, extracted from \eqref{CSols}.
Note that, if $R_\mathrm{L} = R_\mathrm{s} = R^\ast$, we have $\mu < 0$ for all $0 \leq k^2 < 1$, so any $C \geq 2L/(R+R^\ast)^2$ results in a unique positive solution of \eqref{fOmegaEq}. In contrast to the well-known argument in the literature (i.e., to select the squared coupling coefficient $k^2$ to be less than a critical value), we have proven that it is possible to prohibit the frequency splitting phenomenon without compromising or inhibiting the coupling strength.

\begin{figure}[!tbp]
	\centering
	\includegraphics[width=0.45\textwidth]{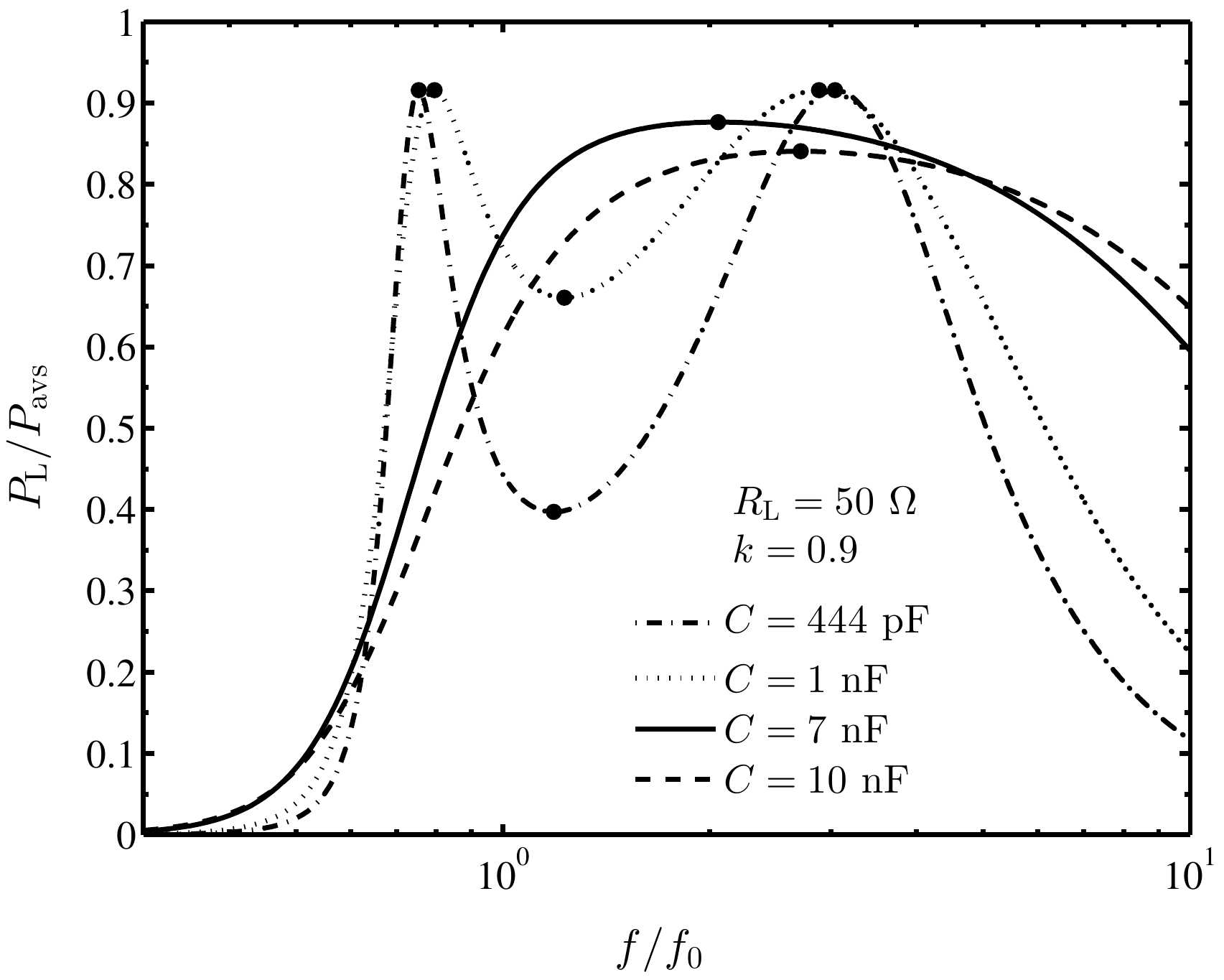}
	\caption{\small Frequency responses of the transferred power with a coupling coefficient of $k=0.9$, a load resistance of $R_\mathrm{L} = 50 \, \Omega$ and different added capacitances, parameterized by two ratios $P_\mathrm{L}/P_\mathrm{avs}$ and $f/f_0$. Solid circle: maximum/minimum power at extreme frequencies given by \eqref{GenSol1}, \eqref{GenSol2} and \eqref{GenSol3}. The power is computed using \eqref{PL2coilsGen}.}
	\label{Fig:FreqSplEliminated}
\end{figure}

Figure \ref{Fig:FreqSplEliminated} presents an example where the frequency splitting is eliminated by choosing appropriate external capacitances, $C = 7$ nF and $C = 10$ nF. The two resonators are set to be identical and an exemplary set of system parameters is as follows \{$L = 9.13$ $\mu$H, $R = 2.25$ $\Omega$, $R_\mathrm{s} = 50$ $\Omega$\}, adapted from \cite{Assawaworrarit2017}. The load resistance and the coupling coefficient are (arbitrarily) kept fixed at $R_\mathrm{L} = 50 \, \Omega$ and $k = 0.9$. The power available from the source is $P_\mathrm{avs} = \abs{V_\mathrm{s}}^2/(8R_\mathrm{s})$ and the primary resonance frequency of the two coils is $f_0 = \omega_0/(2\pi)$. Since $\mu < 0$, the condition of $C$ such that the frequency splitting phenomenon does not exist is $C \geq C_\gamma \approx 6.7$ nF. The capacitances below that value such as $C = 444$ pF and $C = 1$ nF lead to the frequency splitting behavior. 
In a general trend, higher value of $C$ yields lower resonance frequency and (therefore) lower loaded quality factors, which reduces $\beta_\mathrm{s}$ and increases $\gamma_\mathrm{s}$. Until $\beta_\mathrm{s}$ is non-positive and/or $\gamma_\mathrm{s}$ is non-negative, the frequency splitting phenomenon is completely eliminated.

This finding relaxes the requirement of, for instance, sophisticated designs \cite{Lee2013, Lyu2015, Zhang2017}, nonlinear resonant circuits \cite{Abdelatty2019}, frequency/impedance tuning \cite{Heebl2014}, coupling/mutual inductance adaptation \cite{Zhang2014, Aditya2019} or sensing, feedback and control units \cite{Luo2014}, among others. The proposed approach is simpler and easier to implement while maintaining its effectiveness, compared to the conventional methods developed in previous works.

Up to this point, we are able to assert that all the experimental observations of such a frequency splitting reported in the literature are just one of two possibilities that can be performed by a WPTS. If the transmitter and receiver coils are not identical but the parameters $(L_1, \, R_1)$ and $(L_2, \, R_2)$ are close, the same procedure can be applied to approximate the regions of $C_1$ and/or $C_2$ in which the delivered power has only one stationary point. The general problem with arbitrary values of $(R_{1}, \, R_{2}, \, L_{1}, \, L_{2}, \, C_{1}, \, C_{2})$ is much more complicated and its analytical solution is open for future study. However, a numerical solution is possible and is demonstrated through the following example.

Without loss of generality, we assume that $C_2$ is a dependent variable and is given by $C_2 = L_1 C_1/L_2$ to match the resonance frequencies of the two resonators. The numerical problem is formulated as
\begin{align}
\small
C_\mathrm{m} = \min\{C_1\} \label{GenProbNumSol}
\end{align}
such that there exists a unique positive solution among the four $\{\Omega_1,\, \Omega_2,\, \Omega_3, \, \Omega_4\}$. Instead of using the same transmitter parameters, we now consider the case where $L_1 \!=\! 18.26$ $\mu$H and $R_1 \!=\! 4.50$ $\Omega$, which are double of those used in the previous example. The electrical characteristics of the receiver coil are taken from \cite{Assawaworrarit2017}, \{$L_2 = 8.92$ $\mu$H, $R_2 = 2.5$ $\Omega$\}. Solving \eqref{GenProbNumSol} with $k = 0.9$ and $R_\mathrm{L} = 10$ $\Omega$ yields $C_\mathrm{m} \approx 8.01$ nF, while $C_\mathrm{m} \approx 0.55$ nF with $R_\mathrm{L} = 100$ $\Omega$. The frequency splitting is present with any $C_1 < C_\mathrm{m}$ and is suppressed for all $C_1 \geq C_\mathrm{m}$.

As a general principle, lowering both or either of the two $Q$--factors $(Q_\mathrm{s}, \, Q_\mathrm{L})$ to a certain level could assure the suppression of the frequency splitting property \cite{TruongTIE2020}. Other system parameters can then be designed accordingly based on the required values of $Q_\mathrm{s}$ and $Q_\mathrm{L}$. However, the maximum possible power delivered to the load and the optimum inductive-link efficiency heavily depend on the coil quality factors. Therefore, $Q_\mathrm{s}$ and $Q_\mathrm{L}$ should not be chosen too low. Furthermore, even in circumstances where there is not much room for tuning $C_{1/2}$, for instance, ($L_{1/2}, \, R_{1/2}$) are given and the operating frequency is predetermined, $Q_\mathrm{s}$ ($Q_\mathrm{L}$) can still be adjusted by adding an external resistor in series with the transmitter (receiver).
It is important to note that, if a pair of $(C_1, \, C_2)$ satisfies conditions such that the frequency splitting phenomenon is eliminated for a given load resistance $R_\mathrm{L}$, this conclusion also holds for all $R_\mathrm{X} \geq R_\mathrm{L}$. It is a consequence of the general trend of reducing $Q$--factors since a higher resistance leads to a lower quality factor. Therefore, such behavior can be avoided across a wide range of $R_\mathrm{L}$. In a dynamic operating condition where the electrical load varies from $R_\mathrm{m}>0$ to $R_\mathrm{n} > R_\mathrm{m}$, i.e., $R_\mathrm{L} \in [R_\mathrm{m}, \, R_\mathrm{n}]$, choosing suitable values of $(C_1, \, C_2)$ only for $R_\mathrm{m}$ is simultaneously sufficient for the whole range.

While the proposed method (i.e., increasing capacitance to reduce the resonance frequency and quality factor) can be utilized for various applications, it is perhaps the most suitable for biomedical wearable and implantable devices. In these systems, a lower operating frequency is desired as it allows higher permissible external magnetic flux density that can be applied to the human body, according to IEEE safety standards and regulations \cite{IEEE2006}.

\section{Relations between Resonance ($\omega_\mathrm{r}$) and Extreme ($\omega_\mathrm{e}$) Frequencies} \label{ResFreqsVsExtrFreqs}

The aim of this section is to address the question on the effects of the resonance frequencies $^\mathrm{i}\omega_\mathrm{r}$ on the dynamics of the output power. As we have analyzed in \cite{TruongWPTS2018} and recapped in \cite{TruongTIE2020}, the extreme points (i.e., maxima and minimum) of $P_\mathrm{L}$ are achieved at frequencies $^\mathrm{i}\omega_\mathrm{e}$ that are close to $^\mathrm{i}\omega_\mathrm{r}$. This can lead to a misunderstanding on that the resonance frequencies always cause or relate to the frequency splitting phenomenon. However, we find that this might not be necessarily hold true. One simple counterexample can be found in \cite{TruongTIE2020} and a more in-depth analysis is to be presented.

Taking the case when the two coils are identical as an example, if the following conditions are simultaneously fulfilled, equation $\Im \{ Z_\mathrm{in} \} = 0$ has three positive solutions but the frequency splitting behavior does not occur: (a) $C \in \mathbb{R}_{\mu+/\mu-}$, (b) The loaded quality factor of the receiver coil at the primary resonance frequency $^{0}\omega_\mathrm{r} = \omega_0$ satisfies $Q_0 = \omega_0 L /(R + R_\mathrm{L}) > 1/\sqrt{2}$, or equivalently $C \in \R_\mathrm{q} = (0, \, C_\mathrm{q})$ where $C_\mathrm{q} = 2L/(R+R_\mathrm{L})^2$, and (c) $k^2 > k_\mathrm{r}^2 = \Theta (1-\Theta/4)$ where $\Theta = 1/Q_0^2$. The analytical expressions of all the resonance frequencies $^\mathrm{i}\omega_\mathrm{r}$ were presented in Table 1 of \cite{TruongWPTS2018}.

\textit{In summary, if $C \in \{ \mathbb{R}_{\mu+/\mu-} \cap \R_\mathrm{q} \neq \varnothing \}$ and $k \in (k_\mathrm{r}, 1)$, the system has a unique global maximum output power while exhibiting three resonance frequencies.}

Considering the particular example shown in Figure \ref{Fig:FreqSplEliminated} of Section \ref{FreqsplElim}, we have \{$L = 9.13$ $\mu$H, $R = 2.25$ $\Omega$, $R_\mathrm{s} = 50$ $\Omega$, $k=0.9$\}. With $R_\mathrm{L} = 10 \, \Omega$, due to the inequality $\xi_1 < C_\mathrm{q}$, any $C \in (\xi_1, \, C_\mathrm{q})$ (for instance, $C = 9$ nF) leads to the behavior that is addressed above. In contrast, if $R_\mathrm{L} = 90 \, \Omega$, we get $\xi_1 > C_\mathrm{q}$ and $\mathbb{R}_{\mu+/\mu-} \cap \R_\mathrm{q} = \varnothing$, the system has only one resonance frequency $\omega_0 = 1/\sqrt{LC}$ and $P_\mathrm{L}$ attains a unique peak for all $C > \xi_1$ (such as, $C = 5$ nF). This same dynamics is observed if $R_\mathrm{L} = R_\mathrm{s} = R^\ast$ since the two conditions (a) and (b) become $C > 2L/(R+ R^\ast)^2$ and $C < 2L/(R+ R^\ast)^2$ respectively, which cannot happen at the same time. SPICE simulations are performed in order to substantiate these arguments.

\begin{figure}[!t]
	\centering
	\includegraphics[width=0.45\textwidth]{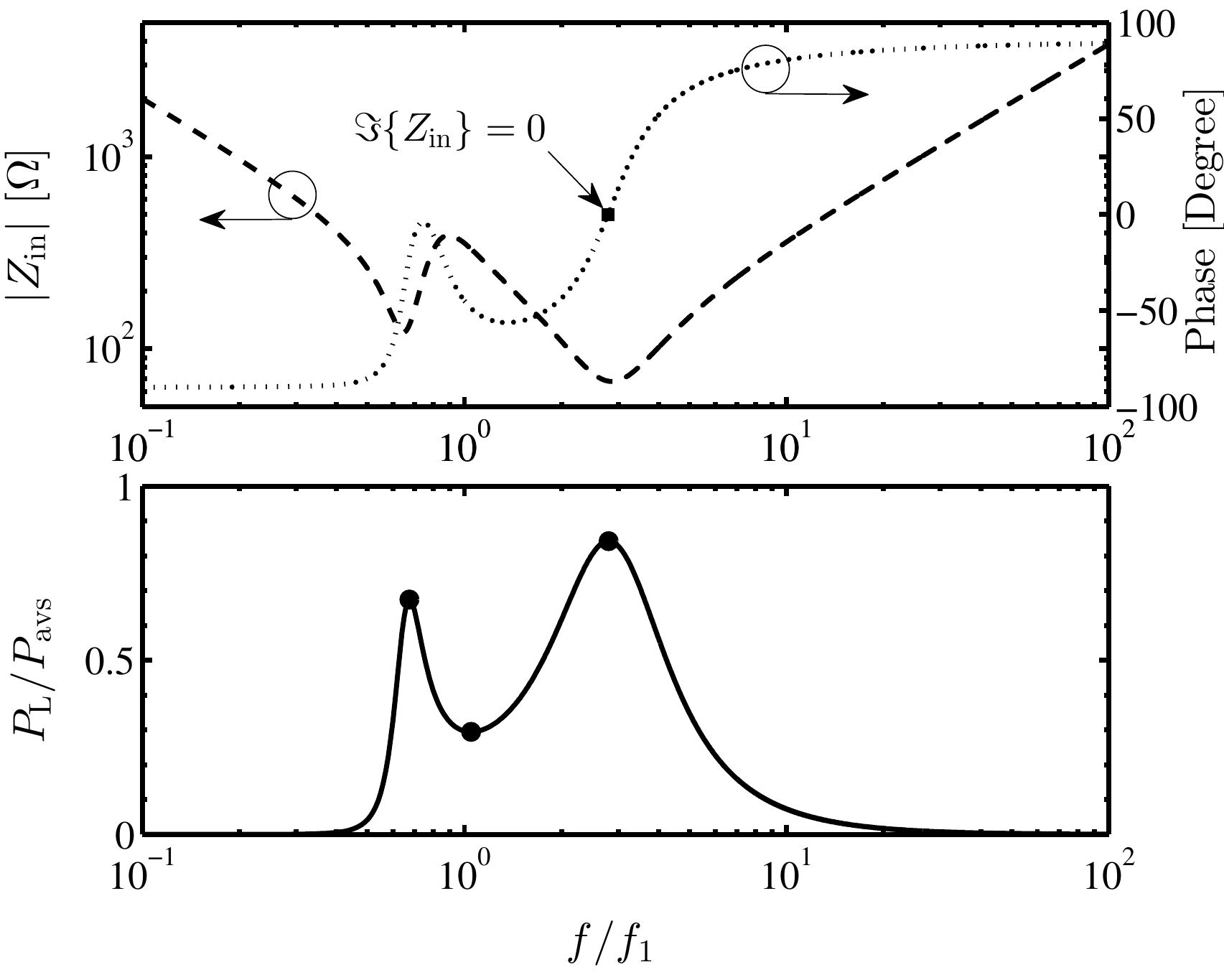}
	\caption{\small Frequency responses of the system input impedance and the power delivered to the load with $k \!=\! 0.9$ and $f_1 \!=\! (2\pi \sqrt{L_1 C_1})^{-1}$. Solid circle: maximum/minimum power at extreme frequencies.}
	\label{Fig:ResVsExtremeFreqs30}
\end{figure}

More interestingly, there also exist circumstances in which equation $\Im \{ Z_\mathrm{in} \} = 0$ exhibits a unique zero-phase angle frequency, however, the frequency splitting phenomenon still takes place. An example is shown in Figure \ref{Fig:ResVsExtremeFreqs30} with an arbitrary system where the two coupled resonators have different resonance frequencies. The parameter set is as follows: \{$L_1 = 18.26$ $\mu$H,
$R_1 = 4.50$ $\Omega$,
$C_1 = 0.444$ nF,
$L_2 = 8.92$ $\mu$H,
$R_2 = 2.50$ $\Omega$,
$C_2 = 1.362$ nF,
$R_\mathrm{s} = 50$ $\Omega$,
$R_\mathrm{L} = 30$ $\Omega$\}.

These results prove that, although analyzing the input impedance and the resonance frequencies might be useful to roughly anticipate the 
frequency splitting behavior in some circumstances, it is not an appropriate approach to correctly comprehend this phenomenon or adequately describe the dynamics of the power transfer to the load. And that, the multiple-resonance and frequency splitting phenomena should be considered independent of each other. This clarification could provide a different perspective on the analysis established in the literature, and prevent any possible further misunderstanding. Dynamic simulations with SPICE are executed to ensure the accuracy of the derivation and verify the results presented in Figures \ref{Fig:FreqSplEliminated} and \ref{Fig:ResVsExtremeFreqs30}.

It is worthwhile to note that by operating the series-series WPTS at either secondary resonance frequencies $^{1}\omega_\mathrm{r}$ or $^{2}\omega_\mathrm{r} \neq\, ^{0}\omega_\mathrm{r}$, a constant output power is obtained regardless of coupling coefficient changes (e.g., due to distance variations or coil misalignment), which is then referred to as coupling-independent transmission regime \cite{Mastri2016}. In general, this constant transferred power is far below the maximum achievable value \cite{Mastri2016}, and therefore utilizing $^{1,2}\omega_\mathrm{r}$ may be not suitable for the purpose of power transfer optimization.

\section{Power optimization and physical bound} \label{PowerOpt}

Although the two-coil WPTS has been widely investigated in the literature, most of the previous work considered the transmission efficiency as the key factor in evaluating the system performance. In this paper, we choose to approach the problem from a different perspective and consider the actual power transferred  to the load as the primary objective of our study. Various aspects of the power optimization are to be investigated.

\subsection{Analytical general-solution to the optimal load}

Determining the optimal load is a powerful method that has been widely used for maximizing the output power of vibration energy harvesters. Since the energy-harvesting concept was proposed as an alternative solution in order to power wireless sensing systems \cite{Roundy2003, Mitcheson2008}, there has been significant effort focusing on two fundamental issues: how much power can be obtained and how to approach it. A method to achieve performance close to the maximum possible output power (that includes displacement-constrained operation) is to adapt the electrical load only \cite{Mitcheson2004b, Truong2016}. 
The two concerns raised above are also solid in the context of WPTS. A closed-form analytical solution for the optimum load that achieves the maximum possible \textit{power efficiency} under arbitrary input impedance conditions was presented in \cite{Zargham2012}. However, it was also shown that a different arrangement is necessary for maximizing the output power of a four-coil WPTS \cite{Dionigi2015}. 

Motivated by these observations, we aim to investigate the optimum power with respect to the load using the conventional gradient descent method. The stationary points of $P_\mathrm{L}$ are evaluated by setting the derivative $\dif P_\mathrm{L}/\dif R_\mathrm{L}$ equal to zero, resulting in
\begin{align}
\small
^\mathrm{opt}\tau \!=\! \frac{L_2}{^\mathrm{opt}R_\mathrm{L}} \!=\! \bigg[ \frac{m_6 \omega^6 + m_4 \omega^4 + m_2 \omega^2}{n_8 \omega^8 \! + \! n_6 \omega^6 \! + \! n_4 \omega^4 \! + \! n_2 \omega^2 \! + \! n_0} \displaystyle \bigg]^{1/2} \label{RLOpt} 
\end{align}
where $\tau_1 = L_1/R_1, \,\tau_2 = L_2/R_2$,
\begin{align*}
\small
m_2 &= \omega_1^4, \, 
m_4 = \tau_\mathrm{s}^{-1} - 2 \omega_1^2 , \,
m_6 = 1, \\
n_0 &= \big(\omega_1 \omega_2 \big)^4 , \\
n_2 &= \big(\omega_2^2/\tau_\mathrm{s}\big)^2 + \big(\omega_1^2/\tau_2\big)^2 - 2 \big(\omega_1 \omega_2 \big)^2 \big(\omega_1^2 + \omega_2^2 \big) , \\
\begin{split}
n_4 &= \tau_\mathrm{s}^{-2} \big( \tau_2^{-2} - 2 \omega_2^2 \big) + \big(\omega_1^2 + \omega_2^2 \big)^2 \\
&- 2 \omega_1^2 \big[ \tau_2^{-2} + (k^2 - 1) \omega_2^2 \big], 
\end{split} \\
\begin{split}
n_6 &= \big(\tau_2^{-1} + \tau_\mathrm{s}^{-1} \big)^2 \\
&+ 2 (k^2 - 1) \big[ (\tau_2 \tau_\mathrm{s})^{-1} + \big(\omega_1^2 + \omega_2^2 \big) \big] ,
\end{split} \\
n_8 &= (k^2 - 1)^2.
\end{align*}
Formula \eqref{RLOpt} is the general closed form expression of the optimal load $^\mathrm{opt} R_\mathrm{L}$ derived as a function of the other system parameters.

\begin{figure}[!t]
	\begin{subfigure}[b]{\linewidth}
		\centering
		\includegraphics[width=0.9\textwidth]{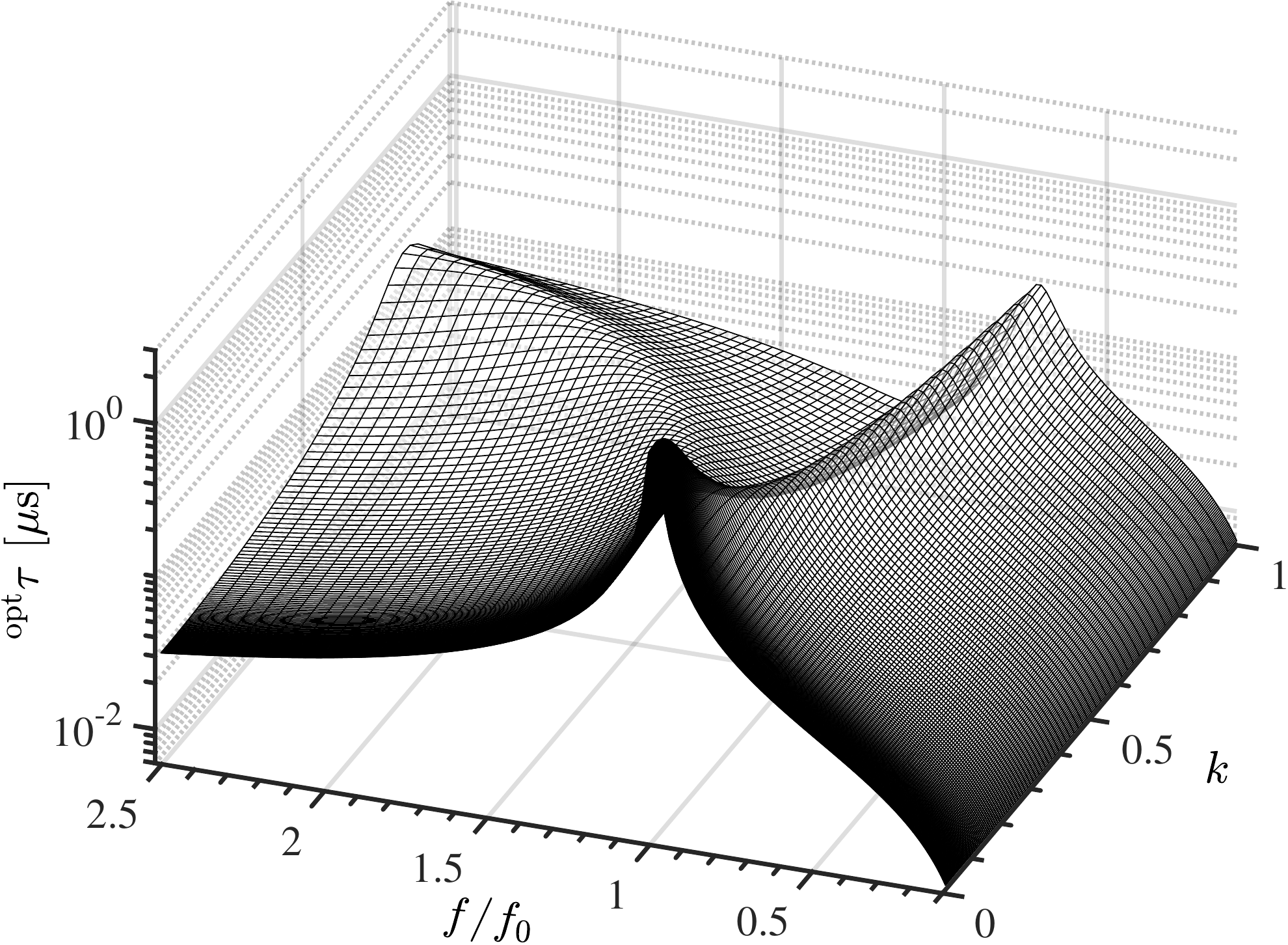}
		\caption{$^\mathrm{opt}\tau$.}
		\label{Fig:OptLoad}
	\end{subfigure}
	\hspace{0.02\linewidth}
	\begin{subfigure}[b]{\linewidth}
		\centering
		\includegraphics[width=0.9\textwidth]{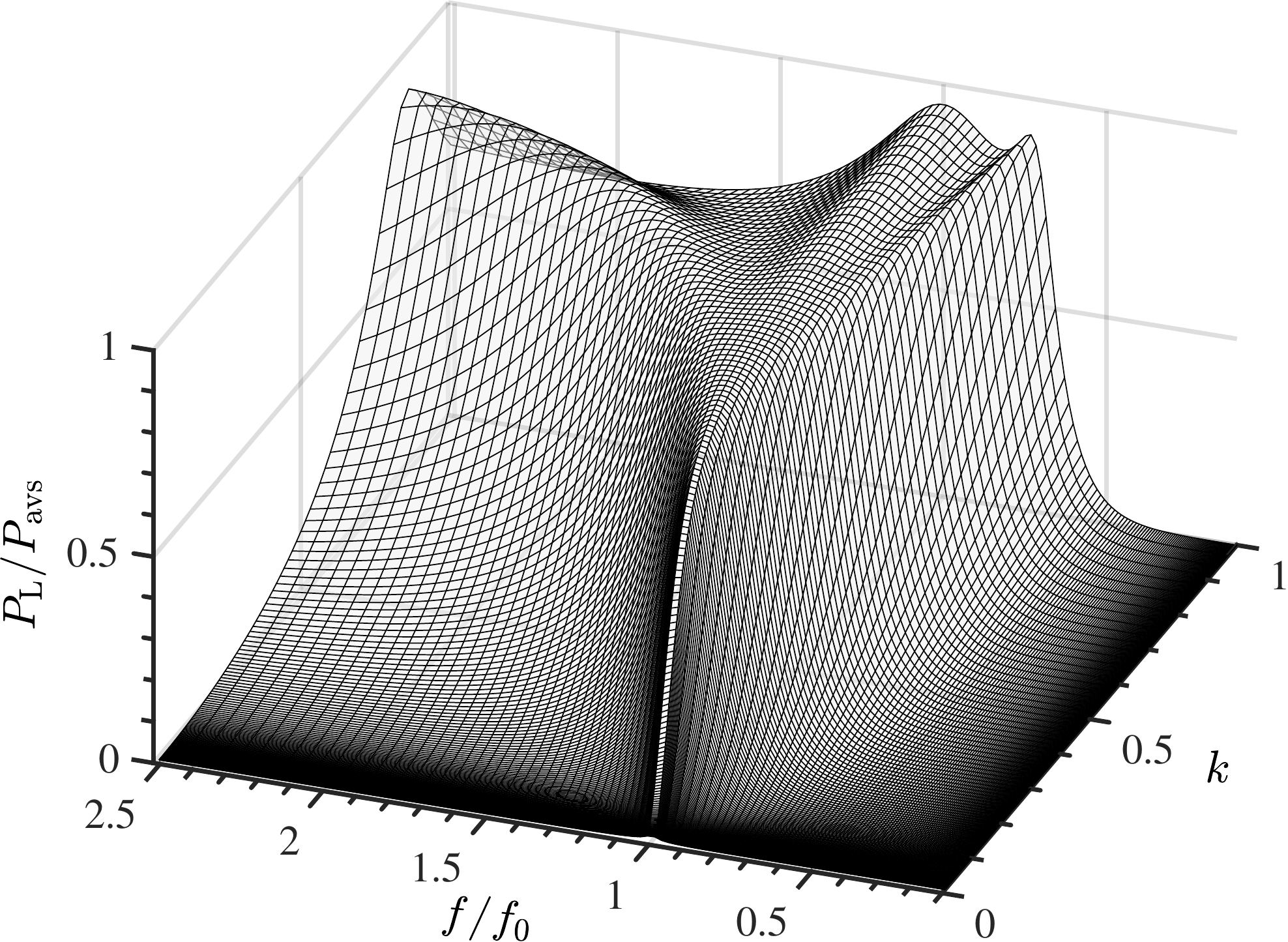}
		\caption{Corresponding optimum output power.}
		\label{Fig:OptPower}
	\end{subfigure}
	\caption{\small Optimal load, characterized by the electrical time constant $^\mathrm{opt}\tau$, and optimum output power $P_\mathrm{L} (k,\, \omega, \, ^\mathrm{opt}\tau (k,\, \omega))$.}
	\label{Fig:OptLoadPower}
\end{figure}
Figure \ref{Fig:OptLoad} shows the changes of $^\mathrm{opt} R_\mathrm{L}$ with respect to $(k,\, \omega)$ characterized by the optimal electrical time scale $^\mathrm{opt}\tau = L_2/^\mathrm{opt} R_\mathrm{L}$ and the corresponding output power as the function of $(k,\, \omega, \, ^\mathrm{opt} R_\mathrm{L})$. Other parameters are taken from the example presented in Figure \ref{Fig:FreqSplEliminated}.
In a general trend, increasing the coupling strength results in decrease of $n_8$ and, therefore, lower (higher) values of $^\mathrm{opt}R_\mathrm{L}$ ($^\mathrm{opt}\tau$). Meanwhile, changes of $^\mathrm{opt} R_\mathrm{L}$ with $\omega$ are much more complicated. Note that, the dynamics of the perfect coupled system where $k = 1$ that makes $n_8 = 0$ is a different scenario and is considered in the next Section. Optimizing the load produces multiple local optimum solutions of the output power with three distinguished branches as shown in Figure \ref{Fig:OptPower}. The optimal load at or nearby the resonance frequency $(f/f_0 \approx 1)$ is perhaps the most convenient option in practice. This observation is discussed in detail in the next Section. Formula \eqref{RLOpt} can also provide a means for designing an optimized WPTS. Given an equivalent load resistance and a transmission distance, \eqref{RLOpt} allows us to estimate appropriate(s) range or value(s) of key parameters such as the transmitter/receiver coil inductance and the selection of a suitable operating frequency. See \cite{Monti2016} as an example of rigorous design for a WPTS where the input impedance, the electrical load and the resonance frequency are specified.

\subsection{Limitations of transferred power} \label{PLboundIdeal}

\begin{figure}[!t] %[!thb]
	\centering
	\includegraphics[width=0.3\textwidth]{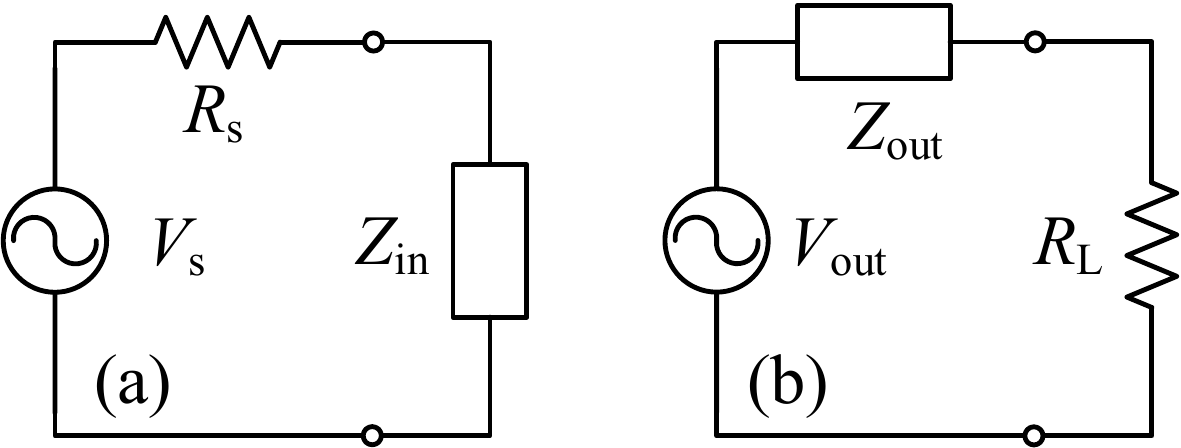}
	\caption{\small Th$\acute{e}$venin equivalent circuits.}
	\label{Fig:Thevenin}
\end{figure}

In order to evaluate the maximum achievable power for a given system, we first consider the conditions for optimizing the power delivered to a load from the two resonators, based on the general theorem presented in \cite{Kong1995}. Here, we consider that the source and load resistances ($R_\mathrm{s}$ and $R_\mathrm{L}$ respectively) are objective variables. A similar technique was used for optimizing the power or OPG of multiple receiver systems in \cite{Fu2015, Monti2017}. The model shown in Figure \ref{Fig:2coilSys} is equivalent to a simple AC circuit in Figure \ref{Fig:Thevenin}a, where $V_\mathrm{out}$ is the Th$\acute{e}$venin equivalent voltage and the output impedance $Z_\mathrm{out}$ is given by \eqref{Z_out}.
The conditions for maximum power are summarized as follows
\begin{align}
\small
\begin{split}
\Im\{Z_\mathrm{out}\} &= \omega L_2 - 1/(\omega C_2) \\
&- \kappa (\omega M)^2 \big( \omega L_1 - 1/(\omega C_1) \big) = 0,
\end{split} \\
\Re\{Z_\mathrm{out}\} &= R_2 + \kappa (\omega M)^2 (R_1 + R_\mathrm{s}) = R_\mathrm{L} \label{RLoptTheorem}
\end{align}
where $\kappa$ is defined by \eqref{kappa} and $\kappa > 0$. One of practical solutions is to adjust the added capacitor $C_2$ on the receiving side and keep other parameters fixed. $C_2$ is then computed by
\begin{align}
\small
C_2 = \frac{1}{\omega} \Big[\omega L_2 - \kappa (\omega M)^2 \Big( \omega L_1 - \frac{1}{\omega C_1} \Big) \Big]^{-1}. \label{C2LoptTheorem}
\end{align}
Substitute the optimal load and secondary added capacitance in \eqref{RLoptTheorem} and \eqref{C2LoptTheorem} back into \eqref{PL2coilsGen}, the transferred power is derived as
%\begin{align}
%\small
%P_\mathrm{L} = \frac{1}{8} \frac{\abs{V_\mathrm{s}}^2}{(R_1 + R_\mathrm{s}) + \displaystyle \frac{R_2}{\kappa (\omega M)^2}}. \label{PLopt_RLZout}
%\end{align}
\begin{align}
\small
P_\mathrm{L} = \frac{1}{8} \frac{\abs{V_\mathrm{s}}^2}{(R_1 + R_\mathrm{s}) + R_2/\big(\kappa (\omega M)^2\big)}. \label{PLopt_RLZout}
\end{align}
Note that $1/\kappa \geq (R_1 + R_\mathrm{s})^2$, therefore
\begin{align}
\small
\begin{split}
P_\mathrm{L} & \leq \frac{1}{8} \frac{\abs{V_\mathrm{s}}^2}{(R_1 + R_\mathrm{s}) \Big[ 1 + R_2 (R_1 + R_\mathrm{s})/(\omega M)^2 \Big]} \\
& < P_\ast = \frac{1}{8} \frac{\abs{V_\mathrm{s}}^2}{R_1 + R_\mathrm{s}} < P_\mathrm{avs} = \frac{1}{8} \frac{\abs{V_\mathrm{s}}^2}{R_\mathrm{s}}. \label{PLbound}
\end{split}
\end{align}
The equality on the left side of \eqref{PLbound} can be attained if and only if $\omega = \omega_1 = 1/\sqrt{L_1 C_1}$. The power available from the source $P_\mathrm{avs}$ is a strict upper bound, however, it does not describe the system performance comprehensively. We further maximize the output power attained in \eqref{PLopt_RLZout} with respect to the driving frequency. The stationary point is determined by setting $\dif P_\mathrm{L}/\dif \omega = 0$, which yields
\begin{align*}
\small
^\mathrm{opt}\omega = \frac{1}{\sqrt{L_1 C_1 - \displaystyle  \frac{1}{2} \big[C_1(R_1 + R_\mathrm{s}) \big]^2}} = \frac{\omega_1}{\sqrt{1 - \displaystyle \frac{1}{2 Q_\mathrm{s}^2}}}
\end{align*}
where $Q_\mathrm{s}$ is defined by \eqref{Qs_at_omega1}. Since the desired self-resonance frequency of the transmitter coil $f_1 = \omega_1/2\pi$ is in the range of MHz, the quality factor $Q_\mathrm{s}$ is significantly larger than unity and the optimal frequency $^\mathrm{opt}\omega$ can be approximated by $\omega_1$. Hence, $P_\mathrm{L} \at[\big]{\omega_1}$ is asymptotic to $P_\mathrm{L} \at[\big]{^\mathrm{opt}\omega}$ and is utilized in following derivations as an alternative for the sake of simplicity. When $\omega = \omega_1$, from \eqref{RLoptTheorem} and \eqref{C2LoptTheorem}, we get 
%\begin{align}
%\small
%R_\mathrm{L} &= R_2 + \frac{(\omega_0 M)^2}{R_1 + R_\mathrm{s}}, \label{RLopt} \\
%\omega_1 &= \frac{1}{\sqrt{L_2 C_2}} = \omega_2. \label{ResMatch} 
%\end{align}
\begin{align}
\small
R_\mathrm{L} &= R_2 + (\omega_0 M)^2 \big{/} (R_1 + R_\mathrm{s}), \label{RLopt} \\
\omega_1 &= 1 \big{/} \sqrt{L_2 C_2} = \omega_2. \label{ResMatch} 
\end{align}

Denoting $\omega_1 = \omega_2$ by $\omega_0$, the output power $P_\mathrm{L} \at[\big]{\omega_0}$ reads as
\begin{align}
\small
\begin{split}
P_0 &= \frac{1}{8} \frac{\abs{V_\mathrm{s}}^2}{(R_1 + R_\mathrm{s}) \big[ 1 + R_2 (R_1 + R_\mathrm{s}) \big{/}(\omega_0 M)^2 \big]} \\
&= \frac{1}{8} \frac{\abs{V_\mathrm{s}}^2}{(R_1 + R_\mathrm{s}) \big[ 1 + 1/(k^2 Q_2 Q_\mathrm{s}) \big]} \\
&= \frac{1}{8} \frac{\abs{V_\mathrm{s}}^2}{R_\mathrm{s}} \Big(1 - \frac{Q_\mathrm{s}}{Q_1} \Big) \frac{k^2 Q_2 Q_\mathrm{s}}{1 + k^2 Q_2 Q_\mathrm{s}} \label{PLmax}
\end{split}
\end{align}
where the unloaded quality factors of the transmitter and receiver coils at the resonance frequency $\omega_0$ are defined by
%\begin{align}
%\small
%Q_1 = \frac{\omega_0 L_1}{R_1}, \,
%Q_2 = \frac{\omega_0 L_2}{R_2}
%\end{align}
\begin{align}
\small
Q_1 = \omega_0 L_1/R_1, \,
Q_2 = \omega_0 L_2/R_2
\end{align}
and $Q_1 > Q_\mathrm{s} = \omega_0 L_1/(R_1 + R_\mathrm{s})$. For given coil parameters and a given distance between the two coils (i.e., $L_1, \, R_1, \, L_2, \, R_2$ and $k$ are known), the first expression in \eqref{PLmax} shows that higher resonance frequency $\omega_0$ results in higher maximum output power. In practice, this can be executed by reducing the added capacitances $C_1$ and $C_2$. In the case when only the load resistance is adapted as in \eqref{RLopt}, the relative relationship of $Q_\mathrm{s}$ and $Q_1$ plays a critical role. Once the ratio $Q_\mathrm{s}/Q_1$ approaches unity, $P_0$ is dramatically reduced. A smaller source resistance is preferable in general. However, for this particular circumstance, if $R_\mathrm{s} \ll R_1$, the ratio $P_0/P_\mathrm{avs}$ is negligibly small, meaning that a tiny amount of power from the source is delivered to the load. For a typical WPTS described in previous examples, the inductances $(L_1, \, L_2)$, the capacitances $(C_1, \, C_2)$ and the resistances $(R_1, \, R_2)$ are in the range of $\mu$H, nF/pF and (a few) $\Omega$ respectively, therefore $Q_2 Q_\mathrm{s} \gg 1$. In the strong coupling regime, $P_0 \xrightarrow{k \to 1} P_\ast$, and if $Q_1$ is much larger than $Q_\mathrm{s}$, then $P_0 \rightarrow P_\mathrm{avs}$.
%%From Table 1: Q_2 Q_\mathrm{s} = 153.8
%\begin{figure}[!b] %[!thb]
%	\centering
%	\includegraphics[width=0.145\textwidth]{TheveninZinEqui.pdf}
%	\caption{\small Equivalent circuit of the WPTS.}
%	\label{Fig:TheveninZin}
%\end{figure}

While some other authors only accounted for the roles of the input power and OPG \cite{Fu2015, Monti2017}, we also further investigate the conditions to maximize the power input to the network from a given source. Similarly, the model in Figure \ref{Fig:2coilSys} is equivalent to the circuit depicted in Figure \ref{Fig:Thevenin}b, where the input impedance $Z_\mathrm{in}$ is defined as \eqref{Z_in}.
Due to \eqref{ResMatch}, the imaginary part of the input impedance is eliminated, $\Im\{Z_\mathrm{in}\} = 0$. The other condition for maximum delivered power is $\Re\{Z_\mathrm{in}\} = R_\mathrm{s}$, which leads to 
%\begin{align}
%\small
%R_\mathrm{s} &= R_1 + \frac{(\omega_0 M)^2}{R_2 + R_\mathrm{L}}. \label{Rsopt} 
%\end{align}
\begin{align}
\small
R_\mathrm{s} &= R_1 + (\omega_0 M)^2 \big{/} (R_2 + R_\mathrm{L}). \label{Rsopt} 
\end{align}

In order to obtain the maximum transferable power from the source, two conditions \eqref{RLopt} and \eqref{Rsopt} need to be satisfied simultaneously. Considering other parameters as dependent variables, the solutions of $R_\mathrm{s}$ and $R_\mathrm{L}$ are expressed as
%\begin{align}
%\small
%R_\mathrm{s} &= R_1 \displaystyle \sqrt{1 + \frac{(\omega_0 M)^2}{R_1 R_2}}, \label{RsFin} \\
%R_\mathrm{L} &= R_2 \displaystyle \sqrt{1 + \frac{(\omega_0 M)^2}{R_1 R_2}} = \frac{R_2}{R_1} R_\mathrm{s}. \label{RLFin} 
%\end{align}
\begin{align}
\small
R_\mathrm{s} &= R_1 \sqrt{1 + (\omega_0 M)^2 \big{/} (R_1 R_2)}, \label{RsFin} \\
R_\mathrm{L} &= R_2 \sqrt{1 + (\omega_0 M)^2 \big{/} (R_1 R_2)} = R_\mathrm{s} R_2 / R_1 . \label{RLFin} 
\end{align}
Substituting \eqref{ResMatch}, \eqref{RsFin} and \eqref{RLFin} into \eqref{PL2coilsGen}, we obtain
\begin{align}
\small
\begin{split}
P_\mathrm{A} & \!=\! \frac{1}{8} \frac{\abs{V_\mathrm{s}}^2}{R_\mathrm{s}} \frac{\displaystyle \sqrt{1 \!+\! \frac{(\omega_0 M)^2}{R_1 R_2}} \!-\! 1}{\displaystyle \sqrt{1 \!+\! \frac{(\omega_0 M)^2}{R_1 R_2}} \!+\! 1} \!=\! \frac{1}{8} \frac{\abs{V_\mathrm{s}}^2}{R_\mathrm{s}} \frac{\displaystyle \sqrt{1 \!+\! k^2 Q_1 Q_2} \!-\! 1}{\displaystyle \sqrt{1 \!+\! k^2 Q_1 Q_2} \!+\! 1} \\
&= P_\mathrm{avs} \frac{k^2 Q_1 Q_2}{\displaystyle \big( \sqrt{1 + k^2 Q_1 Q_2} + 1 \big)^2}. \label{AsymPbound}
\end{split}
\end{align}
%where the unloaded quality factors of the receiver coil at the resonance frequency $\omega_0$ is defined as $Q_1 = \omega_0 L_1/R_1$.
Here, $P_\mathrm{A}$ can be (asymptotically) considered as a tight upper bound of the power that can be delivered to the load with a given power source (characterized by $P_\mathrm{avs}$). The result also quantitatively reveals the crucial role of $Q_1$ and $Q_2$, in which both of them in addition to the coupling strength frankly determine the performance of a WPTS. The asymptotic power bound approaches the power available from the source once $k^2 Q_1 Q_2 \gg 1$.

\begin{figure}[!t]
	\centering
	\includegraphics[width=0.45\textwidth]{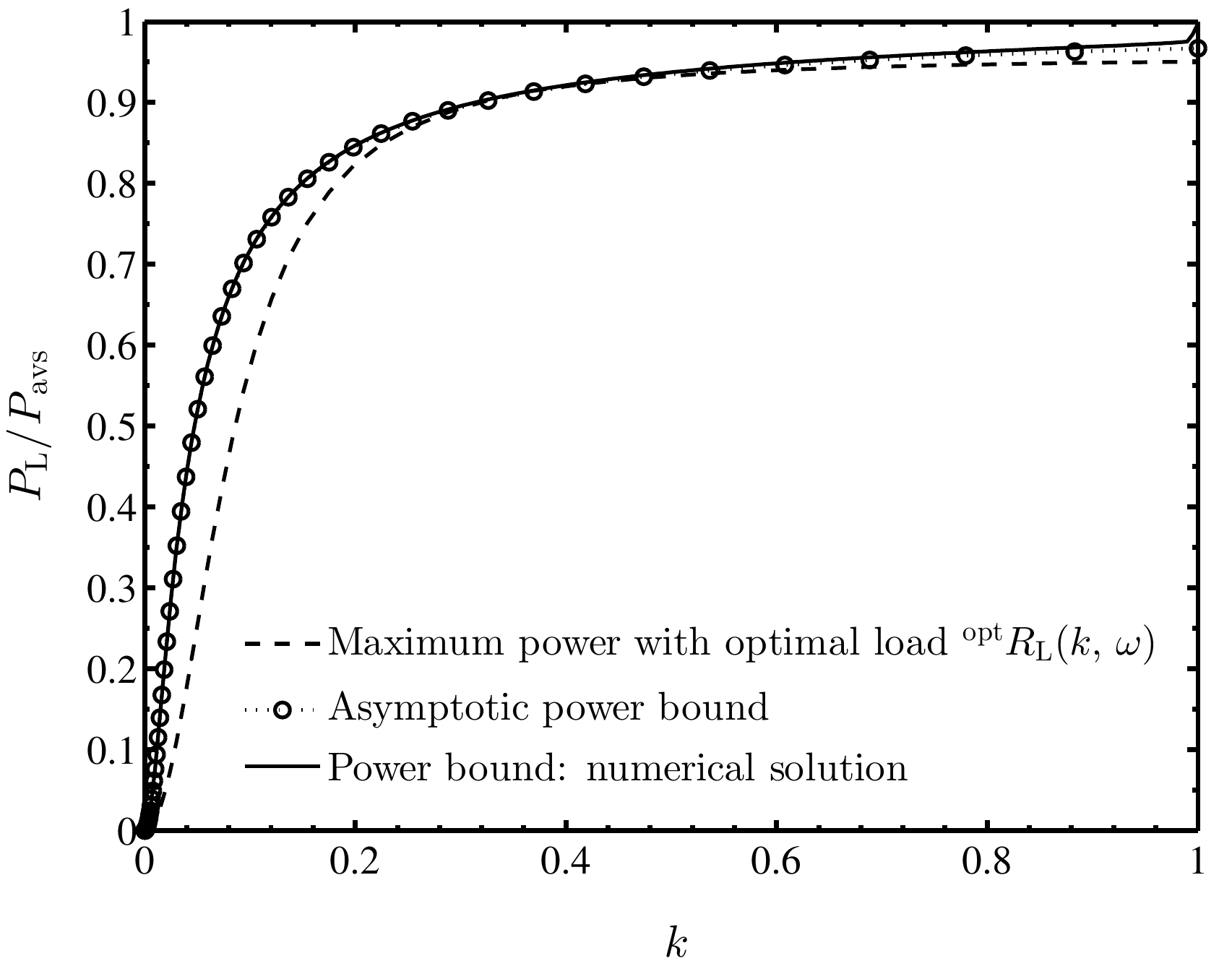}
	\caption{\small Comparison of: (i) The maximum output power presented in Figure \ref{Fig:OptPower} at each coupling strength, (ii) The asymptotic power bound given by \eqref{AsymPbound} and (iii) Numerical solution of the upper power bound \eqref{RigorousBound}.}
	\label{Fig:PowerBound}
\end{figure}

For given resonators where $(R_{1}, \, R_{2}, \, L_{1}, \, L_{2}, \, C_{1}, \, C_{2})$ are know, the rigorous upper bound of the transferable power is determined by
\small
\begin{equation} \label{RigorousBound}
\max_{R_\mathrm{L}, \,R_\mathrm{s}, \, \omega > 0} P_\mathrm{B}
\end{equation}
\normalsize
where $P_\mathrm{B} = P_\mathrm{L}/P_\mathrm{avs}$. We choose to numerically solve this problem due to its high complexity and nonlinearity with inequality constraints of the three variables, using the nonlinear Interior Point (or Sequential Quadratic) Programming methods \cite{Nocedal2006}.
Figure \ref{Fig:PowerBound} presents a comparison of the rigorous upper bound of the output power and the asymptotic closed-form solution. Since they are nearly identical, \eqref{AsymPbound} provides a very convenient and compact method to evaluate the performance of a WPTS. In the case where we are able to arbitrarily adjust the source and load resistances, the maximum transferable power (the physical-bound performance) is attained if three conditions \eqref{ResMatch}, \eqref{RsFin} and \eqref{RLFin} are fulfilled. In addition, the optimum power at each coupling coefficient, $^\mathrm{opt}P_{k}$, extracted from Figure \ref{Fig:OptPower} is included. $^\mathrm{opt}P_{k}$ is verified by SPICE simulations, in which both $\omega$ and $R_\mathrm{L}$ are varied to determine their optimal values at each $k$. In the high coupling regime $(k > 0.2)$, $^\mathrm{opt}P_\mathrm{k} \approx P_\mathrm{A} \approx P_\mathrm{B}$, however there is a significant difference between $^\mathrm{opt}P_\mathrm{k}$ and $\{P_\mathrm{A}, \, P_\mathrm{B}\}$ in the lower regions of $k$. Therefore, only adapting the electrical load at the resonance frequency is sufficient to approach the power bound at strong coupling strength. This can be realized by using asynchronously switched electronic interfaces such as buck-boost converters since they can appear as an effective load resistance determined by the duty cycle of the switching circuit \cite{Lefeuvre2007, Dhulst2010}. Otherwise, adjusting the source impedance is required in a loosely coupled system.

\begin{figure}[!b]
	\centering
	\includegraphics[width=0.49\textwidth]{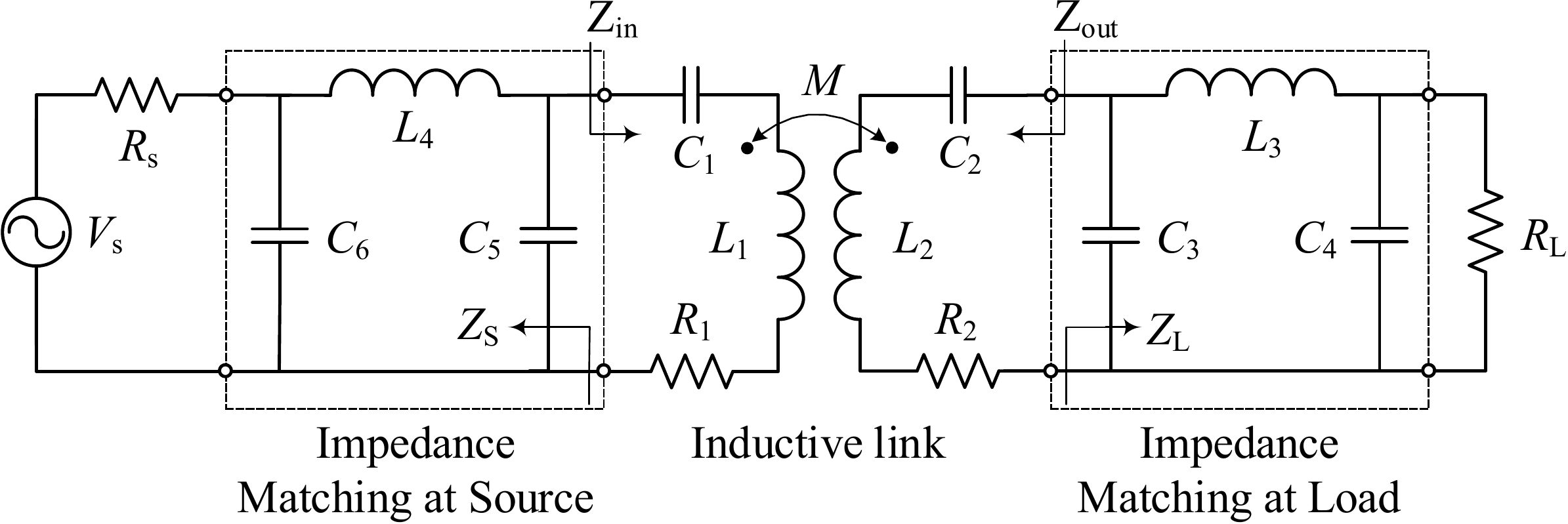}
	\caption{\small Bi-conjugate impedance matching at source and load with two $\Pi-$networks.}
	\label{Fig:PiBiMatch}
\end{figure}

\subsection{Power transfer efficiency: An additional discussion} \label{EffDiscussion}

Dionigi and Costanzo \textit{et. al.} \cite{Dionigi2015, Costanzo2017} reported that maximizing power transfer efficiency (i.e., or operating power gain, defined by the ratio between the generated power and the power input to the network) and maximizing power delivered to the load lead to two different solutions. This raises a question on  the efficiency of the investigated WPTS under the optimum-power operation presented in Section \ref{PLboundIdeal}. 

Two widely used definitions for the power gain of the two-port network are the transducer power gain (TPG), $\eta_\mathrm{t} =$ power delivered to the load/power available from the source, and the operating power gain, $\eta_\mathrm{p} =$ power delivered to the load/power input to the network. $\eta_\mathrm{p}$ is the most common coefficient used in context of wireless power transfer as the transmission (or inductive link) efficiency. At the resonance frequency $\omega = \omega_0$,
\begin{align} 
\small
\eta_\mathrm{t} &= \frac{P_\mathrm{L}}{P_\mathrm{avs}} \at[\bigg]{\omega_0} = \frac{4 R_\mathrm{s} R_\mathrm{L} (\omega_0 M)^2}{\big[(\omega_0 M)^2 + (R_1 + R_\mathrm{s})(R_\mathrm{L} + R_2) \big]^2}, \label{tgain} \\
\eta_\mathrm{p} & \!=\! \frac{P_\mathrm{L}}{P_\mathrm{in}} \at[\bigg]{\omega_0} = \frac{R_\mathrm{L} (\omega_0 M)^2}{\big[(\omega_0 M)^2 + R_1 (R_\mathrm{L} + R_2) \big](R_\mathrm{L} + R_2)}. \label{opgain}
\end{align}
\normalsize
Substituting \eqref{RsFin} and \eqref{RLFin} into \eqref{tgain} and \eqref{opgain}, we get
\begin{align} 
\small
\eta_\mathrm{t} = \eta_\mathrm{p} = \frac{k^2 Q_1 Q_2}{\displaystyle \big( \sqrt{1 + k^2 Q_1 Q_2} + 1 \big)^2}. \label{GlobOpt}
\end{align}
We have revealed that, as the output power reaches its physical bound, $P_\mathrm{in} = P_\mathrm{avs}$ and the two power gains collapse to a unique solution simultaneously. This value is also the maximum possible transmission efficiency of the near-field inductive link \cite{Zargham2012}. Therefore, the global solution of the power optimization problem and that of the efficiency maximization are essentially identical. This interpretation has been extensively studied and validated in the microwave technology society. However, it is still debating in the field of WPT. For example, see \cite{Dionigi2015, Costanzo2017}, in which the authors concluded that optimizing $\eta_\mathrm{t}$ and $\eta_\mathrm{p}$ do not give the same result. This statement is the consequence of the fact that the authors only considered optimizing the load resistance or conjugate matching at the receiver, which results in local optimum point(s). By explicitly deriving equation \eqref{GlobOpt}, we aim to bury the difference between the two optimization schemes and confirm that, as long as we seek for the global maximum, the obtained solutions are equivalent.

Some authors prefer to quantify efficiency instead through the overall/system efficiency,
%\begin{align} 
%\small
%\eta_\mathrm{s} = \frac{P_\mathrm{L}}{P_\mathrm{L} + P_\mathrm{loss}}
%\end{align}
\begin{align} 
\small
\eta_\mathrm{s} = P_\mathrm{L} \big{/} (P_\mathrm{L} + P_\mathrm{loss})
\end{align}
where $P_\mathrm{loss}$ is the total power dissipated in $R_\mathrm{s}, \, R_1$ and $R_2$. Under the optimum-power operation condition, $P_\mathrm{loss} = P_\mathrm{avs} + (P_\mathrm{in} - P_\mathrm{L}) = 2 P_\mathrm{avs} - P_\mathrm{L}$, therefore $\eta_\mathrm{s}$ becomes
\begin{align} 
\small
\eta_\mathrm{s} = \frac{1}{2} \frac{P_\mathrm{L}}{P_\mathrm{avs}} = \frac{1}{2} \frac{k^2 Q_1 Q_2}{\displaystyle \big( \sqrt{1 + k^2 Q_1 Q_2} + 1 \big)^2} < \frac{1}{2}.
\end{align}
This result collides with the general theorem discussed in \cite{Kong1995}, in which the overall efficiency can never go beyond 50 \% as the transferred power is fully optimized.

\subsection{Possible method to obtain maximum output power: bi-conjugate impedance matching circuits}

\begin{figure}[!b]
	\centering
	\includegraphics[width=0.425\textwidth]{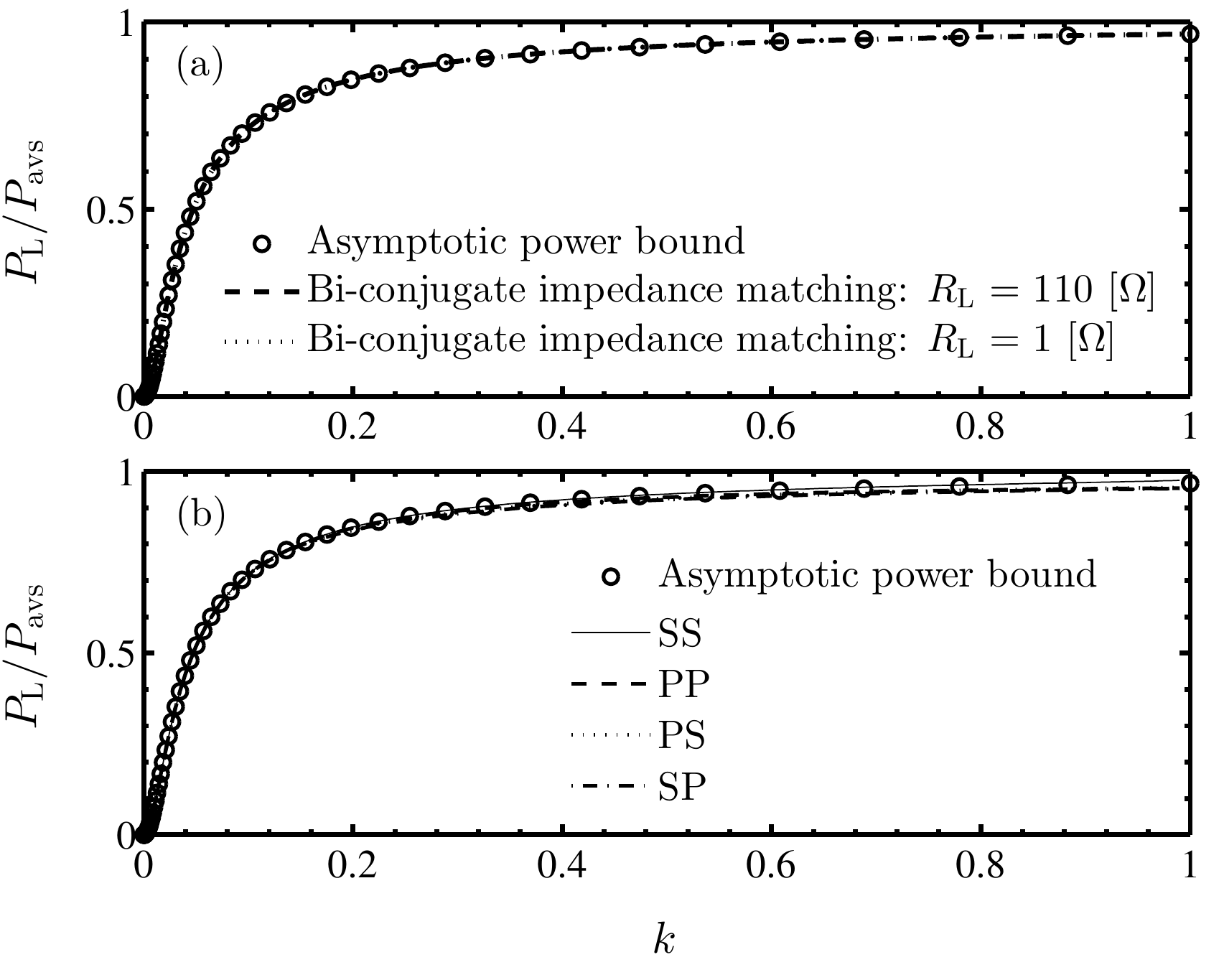}
	\caption{\small (a) Comparison of: (i) The asymptotic power bound given by \eqref{AsymPbound} and (ii) Numerical solutions of the maximum power delivered to two different electrical loads $R_\mathrm{L} = 110 \, \Omega$ and $1 \, \Omega$. (b) Rigorous upper bound of power of four configurations, SS, PP, PS and SP. Notation: S--Series, P--Parallel.}
	\label{Fig:PowerBoundBiMatch}
\end{figure}

In the circumstance where the source, input and load impedances are specified, Monti \textit{et. al.} presented closed-form formulas for the optimum design of a near-field magnetically coupled WPTS, with the main aim is to maximize the transferring efficiency \cite{Monti2016}. The coil inductance $L_\mathrm{i}$, parasitic resistance $R_\mathrm{i}$ and the lumped capacitance $C_\mathrm{i}$ are then determined as functions of desired coil unloaded quality factor $Q_\mathrm{i}$ ($i = \{1, \, 2\}$), given resonance frequency $\omega_0$ and coupling coefficient $k$. This method is probably not appropriate in practice since $L_\mathrm{i}$, $R_\mathrm{i}$ and $k$ are not independent of each other as considered in \cite{Monti2016}.

In realistic application scenarios, it is common that the source impedance and the electronic load are predetermined. This section aims to realize the general power optimization technique presented in Section \ref{PLboundIdeal} under such a situation. Two additional $\Pi-$networks are used as impedance matching circuits at both transmitter and receiver sides, as shown in Figure \ref{Fig:PiBiMatch}. 
The power transferred to the load is derived as follows
\begin{align} 
\small
P_\mathrm{L} = \frac{1}{2} \abs{V_\mathrm{th}}^2 \frac{\abs{Z_{21}}^2 \Re\{Z_\mathrm{L}\}}{\abs{(Z_{11} + Z_\mathrm{s}) (Z_\mathrm{out} + Z_\mathrm{L})}^2} \label{PLWithBiMatch}
\end{align}
where
\begin{align*} 
\small
V_\mathrm{th} &= V_\mathrm{s} \frac{\displaystyle Z_\mathrm{t} \Big[ j \omega C_5 \Big(j \omega L_4 + \big(j \omega C_5 \big)^{-1} \Big) \Big]^{-1}}{\displaystyle R_\mathrm{s} + Z_\mathrm{t}}, \\
Z_\mathrm{t} &= \Big[ j \omega C_6 + \Big(j \omega L_4 + \big(j \omega C_5\big)^{-1} \Big)^{-1} \Big]^{-1}, \\
Z_\mathrm{s} &= \Big[ j \omega C_5 + \Big( j \omega L_4 + \big( j \omega C_6 + R_\mathrm{s}^{-1} \big)^{-1} \Big)^{-1} \Big]^{-1}, \label{ImMatZs_LC}\\
Z_\mathrm{L} &= \Big[ j \omega C_3 + \Big( j \omega L_3 + \big( j \omega C_4 + R_\mathrm{L}^{-1} \big)^{-1} \Big)^{-1} \Big]^{-1}, %\label{ImMatZL_LC} 
\\
\begin{split}
Z_\mathrm{out} &= R_2 + j \big(\omega L_2 - 1/(\omega C_2) \big) \\
& + (\omega M)^2 \Big[ R_1 + \displaystyle j \big(\omega L_1 - 1/(\omega C_1) \big) + Z_\mathrm{s} \Big]^{-1}. %\label{Z_outGen}
\end{split}
\end{align*}
$Z_{11}$ and $Z_{12}$ are calculated as formulas \eqref{Z11} and \eqref{Z12}. Note that \eqref{PLWithBiMatch} has the same explicit form and collapses to \eqref{PL2coilsGen} when replacing $Z_\mathrm{s}, \, Z_\mathrm{L}$ and $V_\mathrm{th}$ by $R_\mathrm{s}, \, R_\mathrm{L}$ and $V_\mathrm{s}$ respectively. 

The bi-conjugate matching conditions can be equivalently formulated as a power optimization problem as
\begin{equation} \label{Eq_OptPL_BiMatchPi}
\small
\max_{C_3, \, L_3, \, L_4, \, C_5 > 0} P_\mathrm{L}
\end{equation}
where two capacitances $C_4$ and $C_6$ are arbitrarily set equal to 10 nF. The driving frequency is $f = f_0$ and the source resistance is $R_\mathrm{s} = 50 \, \Omega$.

Figure \ref{Fig:PowerBoundBiMatch}a shows that the output power obtained by using bi-conjugate impedance matching with two $\Pi-$networks is identical to that resulted from the asymptotic solution, regardless of the load resistance. In other words, formula \eqref{AsymPbound} is able to exactly predict the optimal performance of the series-series WPTS when the two-side impedance matching technique is utilized. The statements presented in Section \ref{EffDiscussion} still holds true under this circumstance. In addition, a few numerical solutions of $C_3, \, L_3, \, L_4$ and $C_5$ are imported to SPICE simulator and the corresponding power outputs are compared to verify the reliability of the optimization formulated in \eqref{Eq_OptPL_BiMatchPi}.
In Figure \ref{Fig:PowerBoundBiMatch}b, we also reveal that the optimum performance of other typical configurations, such as series-parallel (SP), parallel-parallel (PP) and parallel-series (PS) resonators \cite{Li2015}, is close (nearly indistinguishable) to that of series-series structure. Therefore, \eqref{AsymPbound} can be used as an estimated physical power bound for all those three types as well. The power bound of each structure is obtained from solving \eqref{RigorousBound} numerically. Here, Series and Parallel denote the way the coil and added capacitor connect to each other. The question on how to design a particular system (e.g., \cite{Raju2014, Khan2018}) depends on specific applications and is out of scope of the paper.
Although the analysis is applied to a load resistance, all findings in this study are valid for a general complex load (i.e., $Z_\mathrm{L} = R_\mathrm{L} + j X_\mathrm{L}$).

\section{Discussion on dynamics of a perfectly coupled system} \label{UnityCoupling}

We can re-write $P_\mathrm{L}$ in \eqref{PL2coilsGen} as
%\begin{align}
%\small
%P_\mathrm{L} = \frac{a_6 \omega^6}{b_6 \omega^6 + b_4 \omega^4 + b_2 \omega^2 + b_0} \label{PLUnityk}
%\end{align}
\begin{align}
\small
P_\mathrm{L} = a_6 \omega^6 \big{/} (b_8 \omega^8 + b_6 \omega^6 + b_4 \omega^4 + b_2 \omega^2 + b_0) \label{PLUnityk}
\end{align}
where $a_6 \!=\! (C_1 C_2)^2 k^2 L_1 L_2 \abs{V_\mathrm{s}}^2 R_\mathrm{L} \!>\! 0$, $b_8 \!=\! 2 [C_1 L_1 C_2 L_2 (1 \!-\! k^2)]^2 \!>\! 0$, $b_6 \!=\! 2 (C_1 C_2)^2 \big[ L_1(R_2 \!+\! R_\mathrm{L}) \!+\! L_2 (R_1 \!+\! R_\mathrm{s})\big]^2 \!>\! 0$ and $b_0 \!=\! 2$ (other constants can also be extracted from \eqref{PL2coilsGen}, but are not necessary).
Consider an extreme case when the coupling factor between the two resonators is unity, $k = 1$, the coefficient $b_8$ in \eqref{PLUnityk} disappears, $b_8 = 0$.
This leads to a special property of the system, where
%\begin{align}
%\small
%\begin{split}
%\displaystyle ^{\infty}P_\mathrm{L} &= \lim_{\omega \to +\infty} P_\mathrm{L} = \frac{a_6}{b_6} \\
%&= \frac{1}{2} \frac{\abs{V_\mathrm{s}}^2}{R_\mathrm{s}} \frac{\big( \tau_\mathrm{s}^{-1} - \tau_1^{-1} \big) \big( \tau_\mathrm{L}^{-1} - \tau_2^{-1} \big)}{\big( \tau_\mathrm{s}^{-1} + \tau_\mathrm{L}^{-1} \big)^2}  > 0 \label{LimPLUnityk}
%\end{split}
%\end{align}
\begin{align}
\small
\displaystyle ^{\infty}P_\mathrm{L} \!=\! \lim_{\omega \to +\infty} \! P_\mathrm{L} \!=\! \frac{a_6}{b_6}
\!=\! \frac{1}{2} \frac{\abs{V_\mathrm{s}}^2}{R_\mathrm{s}} \frac{\big( \tau_\mathrm{s}^{-1} \!-\! \tau_1^{-1} \big) \big( \tau_\mathrm{L}^{-1} \!-\! \tau_2^{-1} \big)}{\big( \tau_\mathrm{s}^{-1} \!+\! \tau_\mathrm{L}^{-1} \big)^2} > 0 \label{LimPLUnityk}
\end{align}
and is independent of any added capacitances. If the transmitter and receiver are identical, $^{\infty}P_\mathrm{L}$ no longer depends on coil inductances, as follows
%\begin{align}
%\small
%^{\infty}P_\mathrm{L} = \frac{1}{2} \abs{V_\mathrm{s}}^2 \frac{R_\mathrm{L}}{(R_\mathrm{s} + R_\mathrm{L} + 2 R)^2}. \label{LimPLUnitykFinal}
%\end{align}
\begin{align}
\small
^{\infty}P_\mathrm{L} = \abs{V_\mathrm{s}}^2 R_\mathrm{L} \big{/} [2 (R_\mathrm{s} + R_\mathrm{L} + 2 R)^2]. \label{LimPLUnitykFinal}
\end{align}
It should be noted that the coupling coefficient is unity if and only if all the lines of the magnetic flux density of the transmitter coil cuts all of the turns of the receiver coil \cite{Pierce1907}. However, this condition rarely occurs in a finite-dimensional WPTS. Therefore, the considered behavior is difficult to observe in practice.

\section{Experimental validations: Case studies} \label{ExpValidations}

This section is devoted to validating the two following hypotheses. (i) The frequency splitting behavior is not present even at a high coupling strength with the choice of an appropriate external capacitance, and (ii) At a moderate- or strong-coupling regime, the power bound can be approached by optimizing the load resistance at the nominal resonance frequency.

\subsection{Frequency splitting elimination}

An experimental demonstration of frequency splitting suppression was reported \cite{TruongTIE2020}. However, the study presented in this Section aims to provide a more in-depth understanding to complete a comprehensive picture of the equivalent circuit model and the proposed approach. Another goal is to verify an important property in Section \ref{FreqsplElim}, that the quality factors should not be chosen too low, or in other words, the added capacitances should not be too large, in order to keep the transferred power at an appropriate level.

\begin{figure}[!t]
	\centering
	\includegraphics[width=0.4\textwidth]{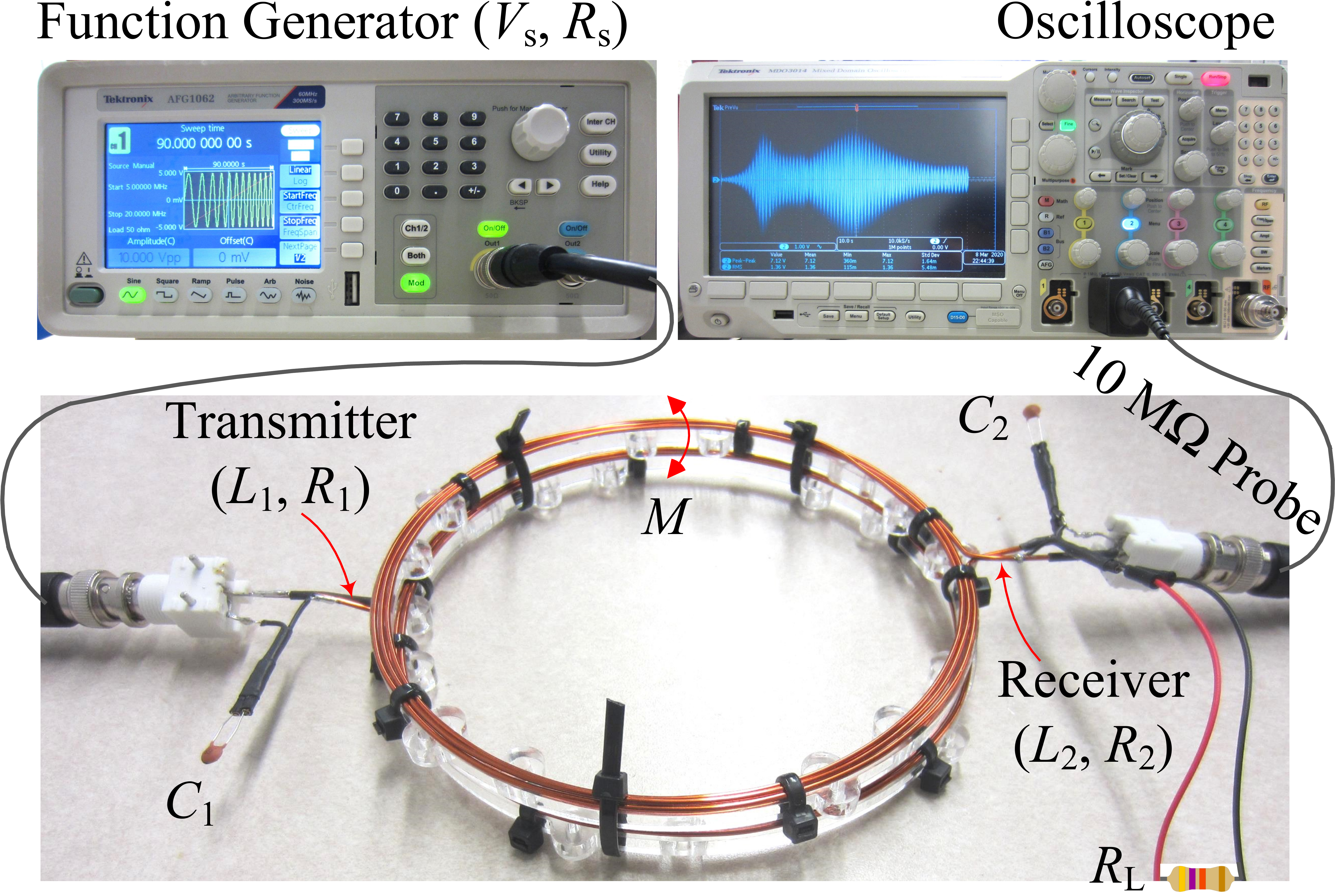}
	\caption{\small Experimental setup of a WPTS in series-series configuration.}
	\label{Fig:ExpSetup}
\end{figure}
Figure \ref{Fig:ExpSetup} shows a schematic illustration of the setup for experiments, in which a Tektronix function generator is simultaneously utilized as a power source and a control unit to drive the transmitter coil. The output voltage induced in the load is measured and recorded by a Tektronix oscilloscope. The source voltage amplitude is set at $V_\mathrm{s} = 10$ V, and the internal source impedance is specified as $R_\mathrm{s} = 50$ $\Omega$. Each transmitter/receiver coil has 3 turns with a diameter of 15 cm and the distance between them is 0.5 cm, approximately. The inductance and resistance of the two coils are $(L_1, \, L_2) = (3.90, \, 3.81)$ $\mu$H and $(R_1, \, R_2) = (0.10, \, 0.71)$ $\Omega$, respectively. The load resistance is arbitrarily chosen as $R_\mathrm{L} = 25$ $\Omega$. In order to analyze the frequency response of the output power, a linearly swept sinusoidal signal is used, with a duration of 90 s. The average power transferred to the load is computed as $P_\mathrm{L} = \frac{1}{t_2 - t_1} \int_{t_1}^{t_2} \frac{(V_\mathrm{L}(t))^2}{R_\mathrm{L}} \dif t$ where $[t_1, \, t_2]$ is the integral time interval.

\begin{figure}[!t]
	\centering
	\includegraphics[width=0.45\textwidth]{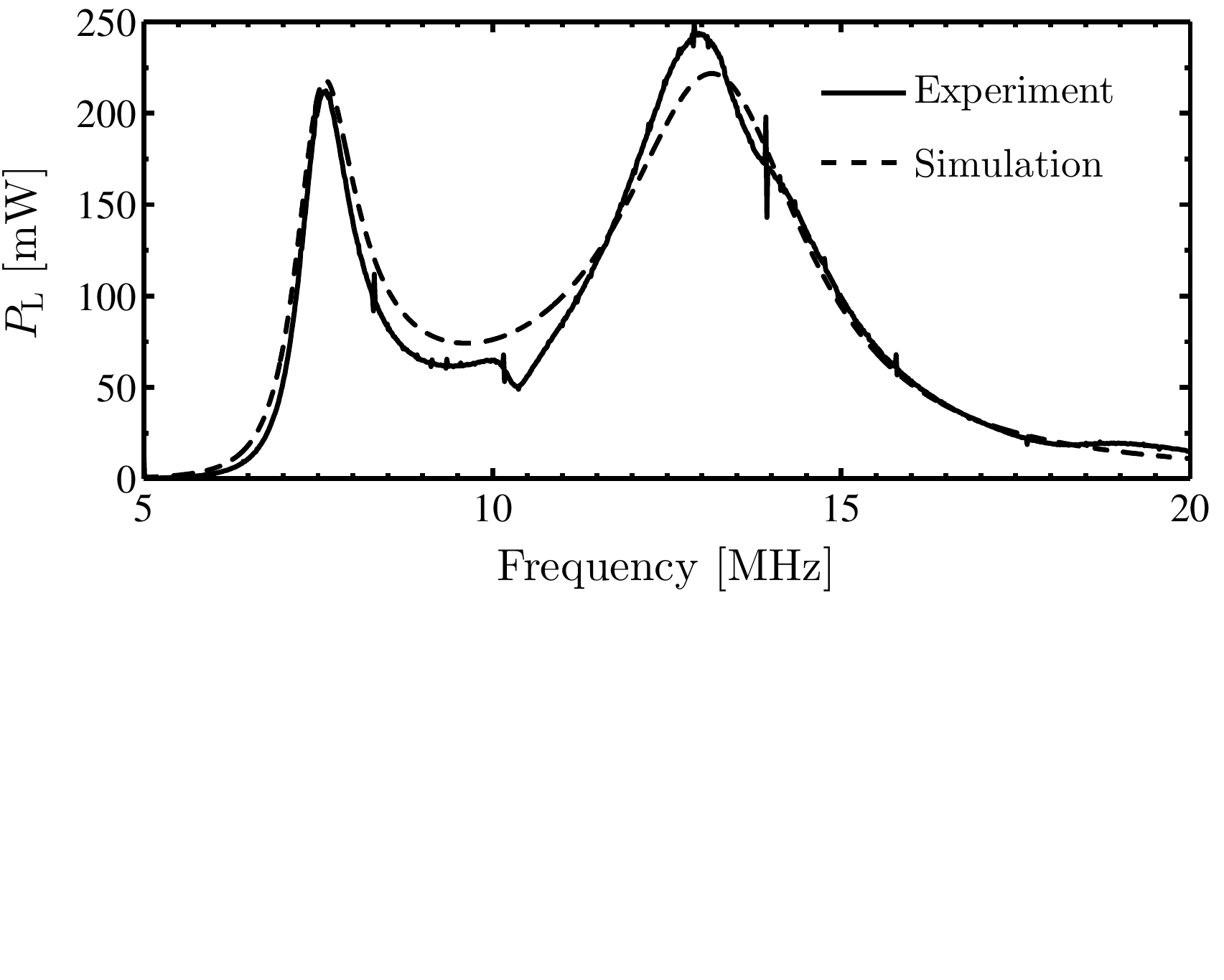}
	\caption{\small Frequency responses of the delivered power with a coupling coefficient of $k = 52.16 \, \times 10^{-2}$, a load resistance of $R_\mathrm{L} = 25$ $\Omega$ and added capacitances of $C_1 = C_2 = 77.08$ pF.}
	\label{Fig:Bifurcation}
\end{figure}
In the first examination, we select a small capacitance of $C_1 = C_2 = 77.08$ pF to ensure the presence of the frequency splitting characteristic, as presented in Figure \ref{Fig:Bifurcation}. Here, the equivalent series resistance (ESR) of these ceramic capacitors are small and neglected. The measured coupling coefficient between the transmitter and receiver is $k = 52.16 \, \times 10^{-2}$ (or $k^2 = 272.06 \times 10^{-3}$); this value is kept fixed while validating all cases. The experimental data and the predictions by the model show a good agreement. 

\begin{figure}[!b]
	\centering
	\includegraphics[width=0.45\textwidth]{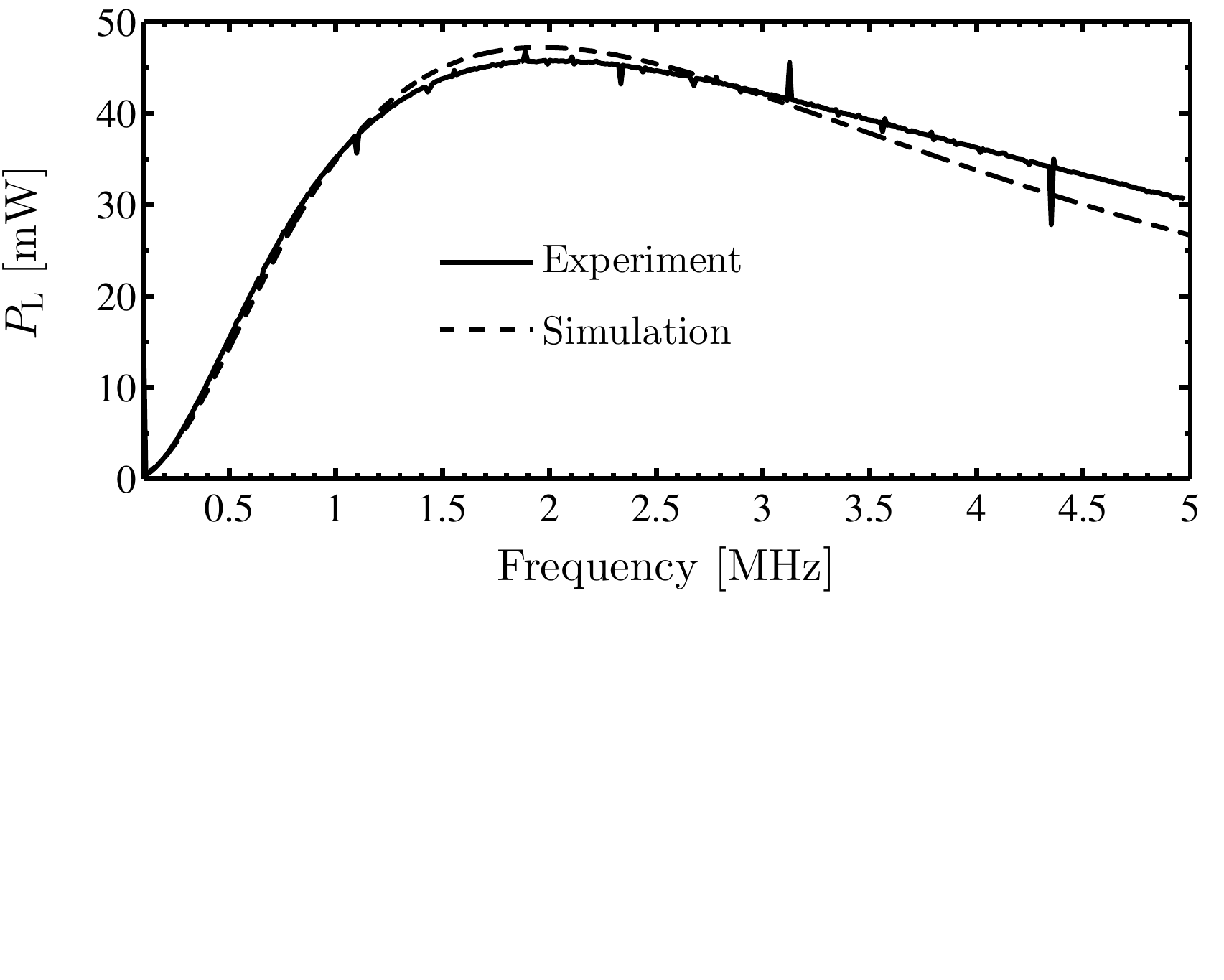}
	\caption{\small Frequency responses of the delivered power with a coupling coefficient of $k = 52.16 \, \times 10^{-2}$, a load resistance of $R_\mathrm{L} = 25$ $\Omega$ and added capacitances of $C_1 = C_2 = 69$ nF.}
	\label{Fig:BifurcationEliminated}
\end{figure}
\begin{figure}[!t]
	\centering
	\includegraphics[width=0.45\textwidth]{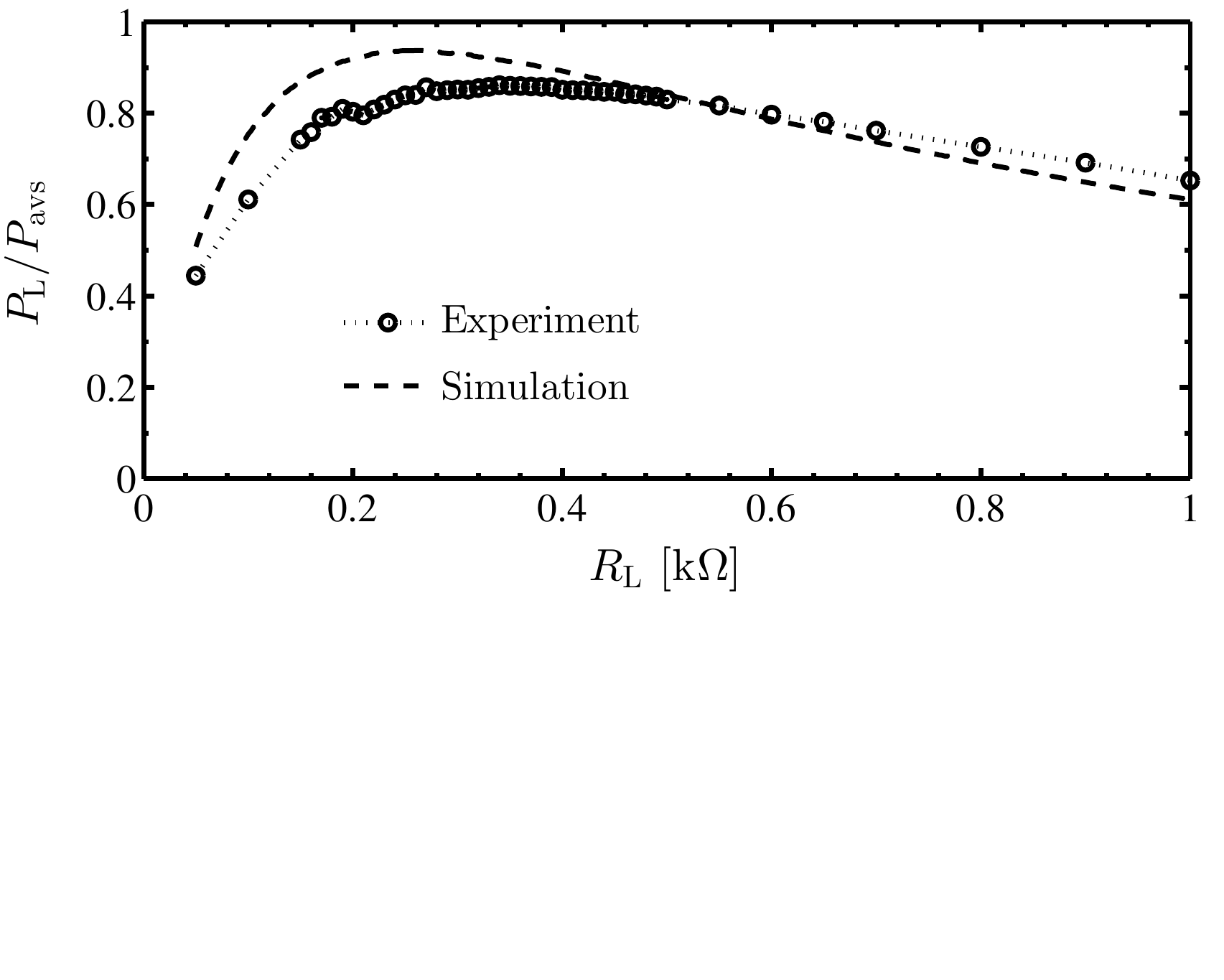}
	\caption{\small Normalized output power (TPG) with respect to load resistance: Comparison between experimental data and simulation results.}
	\label{Fig:PnormRL}
\end{figure}
In order to prove the method proposed in Section \ref{FreqsplElim} is effective and feasible, we now seek a different value of $C_1$ and $C_2$ such that the frequency splitting phenomenon is eliminated, while keeping the other parameters unchanged. For the system under consideration, we have $\mu > 0, \, \xi_1 > 0$ and $C_\gamma < \xi_1$. Here, we use $L \!=\! (L_1 \!+\! L_2)/2$ and $R \!=\! (R_1 \!+\! R_2)/2$ to roughly calculate $\mu, \, \xi, \, C_\gamma$ and other related parameters. Based on Table \ref{Table:C_solution_Sum}, we (arbitrarily) choose $C_1 \!=\! C_2 \!=\! 69$ nF that satisfies the condition $C_{1/2} \in \interval[{C_\gamma, +\infty })$. As a consequence, a single maximum of the power transferred to the load is obtained, as shown in Figure \ref{Fig:BifurcationEliminated}. 
In this case, the two capacitors in use are made of Tantalum and the ERS cannot be negligible. In particular, $R_{C_1} = 4.9 \, \Omega$ and $R_{C_2} = 6.45 \, \Omega$, which are also in series with $R_1$ and $R_2$, respectively. Since we can always take the effects of ERS into consideration by replacing $R_1$ and $R_2$ by effective resistances $R_\mathrm{E_1} = R_1  + R_{C_1}$ and $R_\mathrm{E_2} = R_2  + R_{C_2}$, the disregard of ERS in the original model does not compromise its generality.
Again, the experiment and simulation results are in good agreement. It is important to note that the decrease of $P_\mathrm{L}$ observed in Figure \ref{Fig:BifurcationEliminated} is only a particular case. This behavior is not a universal consequence of the method. As we depicted in Figure \ref{Fig:FreqSplEliminated} and in \cite{TruongTIE2020}, it is possible to select appropriate values of $(C_1, \, C_2)$ so that the frequency splitting property is suppressed while the power transferred to the load is not compromised. 

\subsection{Approaching the power bound at strong coupling}

As the actual transferred power is the central objective of the paper, this subsection aims to verify that, at a strong coupling regime, the power limit can be reached by only tuning the load resistance at (or nearby) the nominal resonance frequency of the two resonators. The setup corresponding to Figure \ref{Fig:Bifurcation} with $C_1 = C_2 = 77.08$ pF is utilized for this purpose.

Figure \ref{Fig:PnormRL} shows the variation of the output power with respect to the load resistance, characterized by the TPG, $\eta_\mathrm{t} = P_\mathrm{L}/P_\mathrm{avs}$. An optimal load of $^\mathrm{opt}R_\mathrm{L} = 0.26$ k$\Omega$ is predicted by \eqref{RLOpt}, while that of the experiment is $0.34$ k$\Omega$; the corresponding maximum TPGs are 0.94 and 0.86. At that given coupling coefficient, the limit of the TPG is $\lim\{\eta_\mathrm{t}\} = 99.5 \times 10^{-2}$, obtained from \eqref{GlobOpt}. The drive frequency is set at $f = 9$ MHz, which is close to the resonance frequencies of the transmitter and receiver ($f_1 = 9.18$ MHz and $f_2 = 9.29$ MHz, respectively). Despite the small discrepancy between the measurements and simulations, the general trend is well captured by the model. This difference can be explained by some factors that are not taken into account, such as the parasitic capacitances of the coils and load, the equivalent series inductance of the added capacitors, or the power loss due to contact resistance.

Finally, the validity of some of the important theoretical analyses in Sections [\ref{FreqSplit}-\ref{PowerOpt}] is justified, including (i) an effective mechanism to eliminate the frequency splitting phenomenon and (ii) approaching the power bound with optimizing the load resistance at moderate/high coupling regimes.

%\vfill

\section{Concluding remarks}

We analyzed the fundamental frequency response and dynamics of a WPTS configured in series, with a focus on the transferred power. We comprehensively revealed the physical insight of the frequency splitting phenomenon by examining the rate of change of the power with respect to the drive frequency rather than considering the characteristics of the input impedance. We also provided a simple but effective method to eliminate this behavior. In particular, high coil quality factors and strong coupling between the two resonators result in the frequency splitting, which can be avoided by adjusting the resonance frequency (or the added capacitors, equivalently). This technique is independent of the coupling strength, meaning that it is not necessary to select the coupling coefficient less than a specific value or limit the minimum distance between the two coils, as suggested by other authors.

The optimal load resistance (i.e., positive stationary point(s) of the equation $\dif P_\mathrm{L}/\dif R_\mathrm{L} = 0$) was derived in a general form and expressed as a function of the other parameters. Numerical computation of the rigorous power bound and analytical asymptotic formula were discussed and compared, showing nearly identical results. The asymptotic solution is therefore an efficient and effective approach to examine the performance of a given WPTS, regardless of configuration (SS, PS, SP or PP). For the completeness of the analysis, a distinctive dynamics of the perfect coupled system, where the coupling coefficient equals unity, is discussed. In this circumstance, it is shown that the output power tends to saturate at a frequency-independent constant value when the driving frequency is relatively large (approaches infinity).

All important findings in this paper were verified by dynamic simulations using SPICE, and in addition, two theoretical analyses were validated with rigorous experiments. Therefore, the mathematical model is reliable and can be used as a framework for designing an optimal system. Although the study is initially motivated by low-power applications, in which the central objective is to optimize the power transferred to the load instead of maximizing the link efficiency, no specific assumptions were made regarding the particular physical nature of the system. Thus, the obtained results and discussion remain valid independently of low- or high-power systems.

\section*{Acknowledgment}
The authors would like to thank the ISS Lab, University of Utah, for providing resources.

\appendices

\section{Tables of signs and variations} \label{SignVariation}

%\medskip
\begin{figure}[!h]
	\small
	%{
	\centering
	\begin{tikzpicture}
	\tkzTabInit[nocadre=false,lgt=1.25,espcl=2]
	{$\Omega$ /1,$\frac{\mathrm{d}P_\mathrm{L}}{\mathrm{d} \omega}$ /1,$P_\mathrm{L}$ /2}{$0$,$\Omega_\mathrm{u}$, $+\infty$}
	\tkzTabLine{,+,$0$,-,}
	\tkzTabVar{-/ $0$ ,+/$P_\mathrm{M-0}$,-/$0$}
	\end{tikzpicture} \par
	%}
	\caption{\small \eqref{FOmegaStand} has unique positive solution.}
	\label{SignVarsUniqueSol}
\end{figure}
%\medskip
%\medskip
%\small
\begin{figure}[!h]
	%{
	\centering
	\begin{tikzpicture}
	\small
	\tkzTabInit[nocadre=false,lgt=1.25,espcl=1.5]
	{$\Omega$ /1,$\frac{\mathrm{d}P_\mathrm{L}}{\mathrm{d} \omega}$ /1,$P_\mathrm{L}$ /2}{$0$,$\Omega_\mathrm{i}$,$ \Omega_\mathrm{j} $, $\Omega_\mathrm{k}$, $+\infty$}
	\tkzTabLine{,+,$0$,-,$0$,+,$0$,-,}
	\tkzTabVar{-/ $0$ ,+/$P_\mathrm{M-1}$,-/$P_\mathrm{m}$,+/$P_\mathrm{M-2}$,-/$0$}
	\end{tikzpicture} \par
	%}
	\caption{\small \eqref{FOmegaStand} has three positive solutions.}
	\label{SignVars3Sols}
\end{figure}
%\normalsize
%\medskip

Figure \ref{SignVarsUniqueSol} shows the signs of $\mathrm{d}P_\mathrm{L}/\mathrm{d} \omega$ and the variations of $P_\mathrm{L}$ when \eqref{FOmegaStand} has unique positive solution $\Omega_\mathrm{u}$. Note that $\alpha < 0$ and $\dif P_\mathrm{L}/\dif\omega \xrightarrow{ \Omega \to +\infty } -\infty$.
The signs of $\mathrm{d}P_\mathrm{L}/\mathrm{d} \omega$ and the variations of $P_\mathrm{L}$ for the case in which \eqref{FOmegaStand} has three positive solutions $0 < \Omega_\mathrm{i} < \Omega_\mathrm{j} < \Omega_\mathrm{k}$ are shown in Figure \ref{SignVars3Sols}. Here $\{i, \, j, \, k\}$ takes one of three values $\{1, \, 2, \, 3\}$ depending on the relations of $\{\Omega_1, \, \Omega_2, \, \Omega_3\}$ in \eqref{GenSol1}, \eqref{GenSol2} and \eqref{GenSol3}. Two maxima and one minimum of $P_\mathrm{L}$ are denoted by $P_\mathrm{M-1}, \,\, P_\mathrm{M-2}$ and $P_\mathrm{m}$ respectively.

\section{Descartes' sign rule} \label{Descartes}
\small
\begin{table}[!hbt]
	\centering
	\caption{All possibilities of Descartes' sign rule for $f(\Omega)$}%
	\begin{tabular}{c c c c c} % centered columns (4 columns)
		%		\hline\hline
		\toprule[1.0pt]
		\centering
		$\alpha$ & $\beta$ & $\gamma$ & $\lambda$ & Maximum number of positive roots, $n$ \\
		\midrule[0.5pt]
		$-$ & $+$ & $-$ & $+$ & 3 \\
		\midrule[0.5pt]
		$-$ & $+$ & $+$ & $+$ & 1 \\
		\midrule[0.5pt]
		$-$ & $-$ & $+$ & $+$ & 1 \\
		\midrule[0.5pt]
		$-$ & $-$ & $-$ & $+$ & 1 \\
		\bottomrule[1.0pt]
	\end{tabular}
	\label{Table:FreqSplit} % is used to refer this table in the text
\end{table}
\normalsize
Table \ref{Table:FreqSplit} shows all possibilities of the signs of $\alpha, \,\, \beta, \,\, \gamma$ and $\lambda$ which are the coefficients of the polynomial $f(\Omega)$. When $k \neq 1$, $\alpha$ is always negative and $\lambda > 0$ for all positive-physical parameters of the system.

%\begin{thebibliography}{1}
%
%\bibitem{IEEEhowto:kopka}
%H.~Kopka and P.~W. Daly, \emph{A Guide to \LaTeX}, 3rd~ed.\hskip 1em plus
%  0.5em minus 0.4em\relax Harlow, England: Addison-Wesley, 1999.
%
%\end{thebibliography}

%\ifCLASSOPTIONcaptionsoff
%\newpage
%\fi

\bibliographystyle{ieeetr} %abbrv

\balance

% that's all folks
\end{document}